%% file: main.tex
\documentclass[10pt,journal]{IEEEtran}

\usepackage{tikz}
\usepackage{pgfplots}
\usepackage{amsmath}
\usepackage{mathtools}
\usepackage{amsthm}
\usepackage{amssymb}
\usepackage{siunitx}
\usepackage{booktabs}
\usepackage{multicol}
\usepackage{multirow}
\usepackage{tabularx}
\usepackage{makecell}   
\usepackage{siunitx}
\usepackage{enumitem}
\usepackage{balance}
\usetikzlibrary{spy}
\usepackage{float}
\usepackage[caption=false,font=footnotesize]{subfig}
\usepackage{pgf-pie}
\usepackage{color,soul}

\usepackage{algorithm}
\usepackage{algpseudocode}
\usepackage{enumitem}

\usepackage{mathtools}

\usepackage{amsthm}
\usepackage{listings}
\usepackage{xcolor}

\definecolor{codegreen}{rgb}{0,0.6,0}
\definecolor{codegray}{rgb}{0.5,0.5,0.5}
\definecolor{codepurple}{rgb}{0.58,0,0.82}
\definecolor{backcolour}{rgb}{0.95,0.95,0.92}
\input{preamble}

\DeclareMathOperator*{\argmin}{arg\,min}

\title{A Reconfigurable and Representation-Adaptive ISA-Based Architecture for Efficient DNN Acceleration}
\author{Vasilis Sakellariou,
         Vassilis Paliouras,
        Ioannis Kouretas,
        Hani Saleh,
        Thanos Stouraitis
\IEEEcompsocitemizethanks{\IEEEcompsocthanksitem Vasilis Sakellariou and Hani Saleh are with the Computer and Information Engineering Department of Khalifa University, Abu Dhabi, UAE
\protect\\
\IEEEcompsocthanksitem Vassilis Paliouras, Thanos Stouraitis, and Ioannis Kouretas are with the Electrical and Computer Engineering Department, University of Patras, Greece }
}

\begin{document}

\maketitle

\begin{abstract}
Domain-specific hardware accelerators provide significantly higher performance and energy efficiency for deep neural network (DNN) workloads than general-purpose processors, but often lack adaptability to evolving model architectures. In contrast, general-purpose ISA-based solutions, such as RISC-V-based accelerators, improve programmability at the cost of efficiency. This work addresses this tradeoff by introducing a machine-learning-oriented instruction set architecture (ISA) and a reconfigurable hardware platform that combine high efficiency with flexibility.
The proposed ISA enables fine-grained control over data movement, dynamic precision, and decoupled execution across data-fetching, tensor processing, and post-processing domains. The corresponding architecture employs lightweight programmable cores and SIMD units to maintain high processing-element utilization with low control overhead, while remaining independent of the underlying numerical representation.
We demonstrate the approach using a Residue Number System (RNS) instantiation supporting $3$--$8$-bit dynamic precision. A 22-nm implementation achieves $5.12$--$10.47$~TOPS/W for a typical workload and up to $1.2\times$  higher energy efficiency than its fixed-point counterpart, while preserving model accuracy. It also outperforms state-of-the-art and mixed-precision accelerators. These results show that the proposed design effectively bridges the gap between efficiency and programmability in modern DNN accelerators.
\end{abstract}


\section{Introduction}

Recent advances in domain-specific hardware accelerators have delivered substantial performance and efficiency gains for AI workloads, particularly in computer vision (e.g., CNNs) and natural language processing (e.g., transformers). State-of-the-art AI accelerators can achieve power efficiencies of $10-50$~TOPS/W~\cite{dnnaccel1,dnnaccel2,dnnaccel3,dnnaccel4,dnnaccel5,dnnaccel6}, marking three orders-of-magnitude improvement over general-purpose processors~\cite{jssc25}. These gains stem from specialized compute units, tightly coupled memory hierarchies, and optimized dataflows tailored to a narrow class of models. However, such accelerators typically rely on fixed memory access patterns, rigid compute pipelines, and bespoke data-movement hardware, which significantly limit their adaptability. As the deep neural network (DNN) landscape evolves rapidly---introducing new model architectures, operators, and optimization techniques---these fixed-function designs often suffer from suboptimal utilization, particularly for workloads that deviate from their target assumptions~\cite{dnnaccel6}. Retargeting these accelerators to new applications requires extensive hardware redesign and verification effort, increasing development cost and deteriorating time-to-market~\cite{datamaestro}.

An orthogonal approach to addressing flexibility has been the adoption of instruction-set-architecture(ISA)-based solutions, such as RISC-V ~\cite{xpulpnn,vega,marsellus,mx,flexv,spatz,quadrilatero} and commercial architectures with vector or matrix extensions (e.g., AMX, Armv8). These systems improve programmability and design reuse by coupling general-purpose cores with tightly integrated accelerator engines (e.g., matrix multiplication and convolution units) via SIMD or vector extensions. While this approach enhances flexibility and provides a programming interface, it introduces area and power overheads associated with general-purpose execution and often struggles to match the performance of specialized ML accelerators~\cite{flexv}.

Beyond architectural organization, numerical data representation is a critical determinant of accelerator efficiency, exposing tradeoffs across area, power, latency, and model accuracy. A wide range of representations, including fixed-point, reduced-precision block floating-point formats, as well as non-conventional representations, have been explored for ML acceleration~\cite{nnsurvey}. However, integration of data representation, quantization, and emerging model optimization techniques within a unified, reconfigurable hardware–software framework remains largely unexplored. Addressing this gap is essential for enabling ML accelerators that combine high efficiency with adaptability across rapidly evolving workloads.

We propose a machine-learning-oriented ISA and a reconfigurable hardware architecture that combine the efficiency of fixed-function accelerators with the flexibility of general ISA-based systems, enabling high compute utilization with minimal control overhead across a wide range of workloads. 
The main contributions of this work are:

\begin{enumerate}
    \item  A custom, extensible instruction set tailored for DNN workloads that enables flexible execution while maintaining high processing-element utilization.
    
    \item  A reconfigurable hardware architecture that efficiently executes the proposed ISA using lightweight programmable cores and dynamic-precision SIMD units. A key feature is the decoupling of control, data movement, tensor computation, and post-processing.
    
    \item A Residue Number System (RNS) implementation supporting $3$--$8$-bit dynamic precision, achieving improved model accuracy–efficiency tradeoffs. The 22-nm placed-and-routed design delivers $5.14$--$10.47$ TOPS/W and up to $1.2\times$ higher energy efficiency than a fixed-point (FXP) counterpart, while preserving model accuracy. It also outperforms state-of-the-art uniform-precision RNS accelerators by up to $1.4\times$, as well as mixed-precision FXP designs at comparable bitwidths.
\end{enumerate}

\section{Proposed Hardware Architecture and ISA}

In contrast to existing general-purpose ISA-based approaches, the proposed ISA is explicitly tailored for ML processing and comprises a minimal set of specialized instructions. The main advantages of the proposed architecture are summarized as follows:
\begin{itemize}
    \item \textbf{Programmability:} Supports a wide range of DNN layers, including convolution, matrix multiplication, attention, skip connections, and depth-wise convolution.
    \item \textbf{Flexible dataflow control:} Enables fine-grained, fully programmable control over data movement and memory addressing.
    \item \textbf{Low control overhead:} Lightweight control/decoding logic and instruction caching minimize area and energy overhead.
    \item \textbf{Dynamic precision:} Supports two types of adaptable precision for improved efficiency across workloads.
    \item \textbf{Rich post-processing:} Provides multiple activation functions and post-processing operations.
    \item \textbf{High PE utilization:} Maintains high processing-element utilization across diverse workloads, due to the decoupling of control and processing domains.
    \item \textbf{Extensibility:} Easy addition of new features (e.g., additional post-processing instructions or memory hierarchy enhancements).
    \item \textbf{Data-representation independence:} Operates independently (at the ISA level) of the underlying numerical representation and can thus enable systematic design-space of data representations under varying accuracy, performance, and hardware cost constraints.
\end{itemize}

\subsection{Processing Domains and Data Streams}

\begin{figure*}[ht!]
    \centering
    \includegraphics[width=\linewidth]{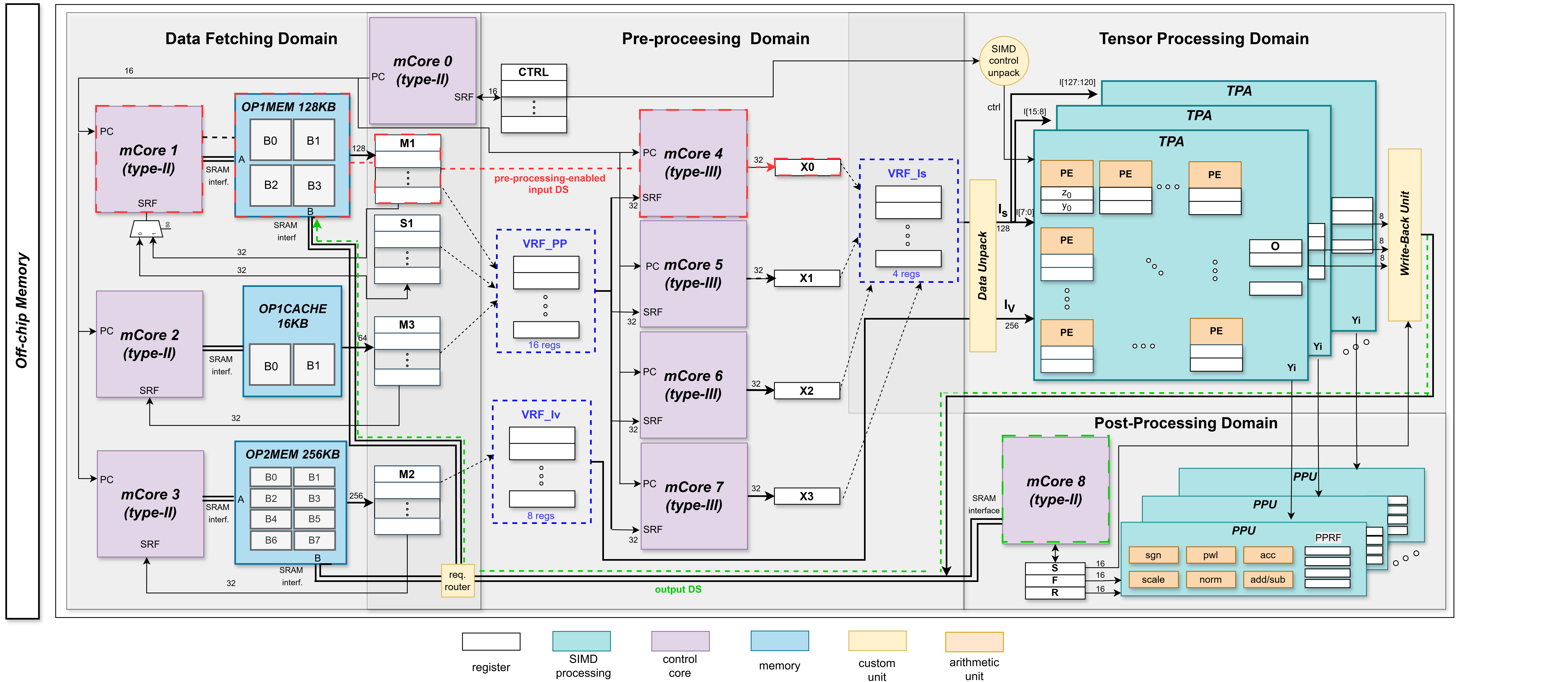}
    \caption{High-level architecture. The proposed processing scheme is organized into four domains: (a) Data-fetching, (b) Pre-processing, (c) Tensor Processing, and (d) Post-processing. Lightweight programmable cores (mCores) control these domains and orchestrate data movement. Structured data streams (DSs) deliver data from the memory hierarchy to the processing units and vice versa (the input DS from OP1MEM to the TPAs is outlined in red). VRFs (outlined in blue)  are virtual groupings of the system's physical registers into a continuous address space, used to facilitate reconfigurability. Each TPA receives a scalar operand \( \mathbf{I_S} \), broadcast to all PEs within the array, and a vector operand \( \mathbf{I_V} \), with one element mapped to each PE. PPUs perform post-processing operation (i.e., activation functions), and results are written back to memory via an output DS (outlined in green).}
    \label{fig:refernce_arch}
\end{figure*}

The proposed architecture uses a \emph{purely integer} SIMD-based processing paradigm that supports both tensor-to-tensor (linear) operations (e.g., matrix multiplication, convolution) and non-linear unary operations (e.g., activation functions). The design comprises multiple programmable control cores, referred to as \emph{mCores}, which execute a program using the developed custom instruction set. The mCores abstract the FSM-based hardwired control logic of conventional accelerators into a lightweight, flexible, and extensible programming model. 

The proposed processing scheme is organized into four distinct processing domains:
(1) the \emph{Data Fetching Domain},
(2) the \emph{Pre-processing Domain},
(3) the \emph{Tensor Processing Domain}, and
(4) the \emph{Post-processing Domain}.
To orchestrate data movement across these domains, the architecture defines a configurable number of \emph{data streams (DSs)}. A data stream represents a structured data flow between the memory hierarchy and the SIMD processing domains, or vice versa. Depending on its role, a DS may carry input activations, model weights, or intermediate feature maps.

An \emph{input data stream} comprises: (i) a memory subsystem where data initially resides; (ii) a control mCore responsible for generating memory addresses and read/write signals; (iii) an optional set of pre-processing mCores that perform transformations on the data; and (iv) a destination register file that stores the final operands supplied to the Tensor Processing domain. An \emph{output data stream} consists of: (i) a source register file that temporarily holds the results produced by the SIMD units; (ii) a memory subsystem; and (iii) a programmable mCore, responsible for committing these results back to the memory hierarchy.

Each data stream is configured through the following parameters:
(a) the number of memory banks (each bank having a fixed 32-bit word width);
(b) the arithmetic precision mode, which may operate in either fine-grained dynamic precision or coarse-grained precision (see Subsection~\ref{s:mxp}); and
(c) a pre-processing enable flag, indicating whether data is forwarded directly to the Tensor Processing Domain or first routed through the Pre-processing Domain.

\subsubsection{Data Fetching Domain}
The \emph{Data Fetching Domain} comprises the memory hierarchy and the programmable mCores that generate memory-control and address signals for each input data stream. The memory hierarchy is organized into multiple independent memory subsystems, each associated with a distinct input DS.
Each memory subsystem is partitioned into multiple memory banks, enabling parallel data access. All subsystems expose a 32-bit global address space. Dedicated address-mapping logic translates the global address into bank-specific local addresses according to a configurable mapping scheme. For example, in a subsystem with two memory banks, the lower 16 bits of the global address may index the first bank while the upper 16 bits index the second. Alternatively, both banks may be indexed using the same subset of global address bits, depending on the desired data layout and access pattern. This address translation logic is customized by the designer when instantiating a variant of the architecture. Upon a valid read operation, the output of each memory bank is latched into a dedicated register. These registers are accessible by the mCore controlling the corresponding memory subsystem and, when pre-processing is enabled, by the mCores in the Pre-processing Domain. 

An input and an output DS may share the same memory subsystem (e.g., for feature-map storage). In this case, the memory must support concurrent read and write accesses, which can be achieved using dual-port or 1-read-1-write (1R1W) memory macros. When native dual-port macros are unavailable, or their additional port complexity is undesirable, we employ pseudo-1R1W constructed from two single-port SRAM macros. The address space is partitioned between the two macros, with the lower and upper halves mapped to separate banks. Concurrent read and write operations are therefore possible as long as the accesses target different halves of the address space.

\subsubsection{Pre-processing Domain}

The \emph{Pre-processing Domain} is positioned between the Data Fetching Domain and the Tensor Processing Domain. Its primary role is to perform lightweight arithmetic and data-manipulation operations on operands retrieved from the memory hierarchy, in cases where rearrangement, alignment, or partial combination is required prior to SIMD execution. Such transformations are commonly required in convolutional layers, where input feature maps may need to be reorganized (e.g., window extraction, padding handling, or stride alignment) before being supplied to the Tensor Processing Domain. The Pre-processing Domain comprises a set of programmable mCores capable of executing a restricted subset of the ISA, sufficient for simple arithmetic and data-movement operations. The domain also includes the register files that temporarily store the transformed operands and serve as the interface to the Tensor Processing Domain.

\subsubsection{Tensor Processing Domain}

The \emph{Tensor Processing Domain} comprises multiple SIMD arrays responsible for executing the elementary arithmetic operations required by DNN workloads.
Each  tensor processing array (TPA) contains \( N_{\text{PE}} \) processing elements (PEs) and three register files $\mathbf{Z, Y}$ and $\mathbf{O}$ of $N_{\text{PE}}$ registers each. The register file $\mathbf{Z}$ is formed by the set of internal PE accumulation registers (used during MAC operations), while $\mathbf{Y}$ stores the final sum when accumulation is complete. Finally, register file $\mathbf{O}$ stores the final result of a DNN operation (i.e., after the application of the activation function). $\mathbf{Z}$ and $\mathbf{Y}$ are extended-precision register files (this precision is reconfigurable), while $\mathbf{O}$ is an 8-bit register file. Details about the functionality of these registers are given in Sections~\ref {s:isa} and \ref{s:templates}. Each TPA receives a scalar operand \( \mathbf{I_S} \), broadcast to all PEs within the array, and a vector operand \( \mathbf{I_V} \), with one element mapped to each PE. While these operands typically correspond to weights and activations, the architecture imposes no semantic constraints on their roles, thereby supporting a broad class of tensor-to-tensor computations.

\subsubsection{Post-processing Domain}

The \emph{Post-processing Domain} consists of multiple post-processing units (PPUs) that also operate in a SIMD fashion. These units are responsible for applying non-linear activation functions. PPUs can also be used to apply element-wise operations between the current output and a previously generated tensor. The activation functions are implemented using piecewise-linear approximations to balance computational efficiency and numerical accuracy. In addition to activation functions, the PPUs support reduction operations, normalization, and sign/overflow detection. By decoupling tensor processing and post-processing into independent domains, the latency of additional functions can be overlapped with ongoing tensor computations, thereby improving overall throughput and resource utilization. PPU micro-architecture is detailed in Section~\ref{s:ppu}.

\subsection{Instruction Set}
\label{s:isa}

The proposed ISA defines 32-bit instructions organized into five categories. The operands and functionality of the basic instructions are summarized in Table~\ref{t:isa}.

\begin{enumerate}
    \item \textbf{Load/Store}: These instructions transfer data between memory and the register files. Load instructions do not specify a destination register, as each memory bank is mapped to a predefined register accessible by the corresponding mCores. Both load and store instructions include a \textit{bank enable} field to select the active memory banks. A variant of the load instruction (\texttt{ld\_add}) performs an additional arithmetic operation on a specified register. A \textit{copy (cp)} flag optionally preserves the previous contents of the destination registers by copying them to an auxiliary register set prior to the read operation. Multiple store modes are supported, as described in Section~\ref{s:wb-unit}.
    Two types of store instructions are defined. Local store (\texttt{st}) stores the contents of a local register to a memory location pointed to by an address register. SIMD store instructions \texttt{st\_simd} collect the contents of TPA output registers $\mathbf{O}$ selected by TPA index $i$ and PE index $j$, pack them according to the \textit{transpose (tr)} and \textit{half precision (hp)} into 32-bit words 
using a dedicated and customizable hardware unit (\textit{write-back unit}), detailed in Section~\ref{s:wb-unit}, and store them in memory. Post-increment/decrement of the address register is also supported in store instructions.
    
    \item \textbf{Arithmetic}: The ISA supports basic logical operations and addition/subtraction (multiplication and division are not included), with both register and immediate operands. Add/sub instructions provide byte-level preshift and masking capabilities (see Table~\ref{t:isa}). Conditional addition (\texttt{addic}) enables efficient implementation of control logic, based on the ALU zero flag ($zf$). The \texttt{ldi} instruction loads a 16-bit immediate value into a register, while \texttt{ldid} replicates the 16-bit constant across both halves of a 32-bit register.
    
    \item \textbf{Branch}: Both conditional and unconditional branching are supported, including combined decrement-and-branch functionality via \texttt{bnzd}.
    
    \item \textbf{SIMD}: This category comprises instructions offloaded to the SIMD processing units, including both the TPAs and the PPUs. SIMD instructions operate on the register sets $(\mathbf{Z}, \mathbf{Y}, \mathbf{O})$ and on the dedicated post-processing register file \textbf{PPR} within each PPU. The \texttt{mac} instruction triggers a multiply--accumulate operation across all PEs, using the two SIMD operands and the local accumulation register  $\mathbf{Z}$. The \texttt{pproc} instruction transfers data from $\mathbf{Z}$ to $\mathbf{Y}$ and generates an interrupt to the post-processing control mCore, enabling decoupled execution. PPU instructions include: (i) activation function instructions (\texttt{afunc}), which apply non-linear functions to $\mathbf{Y}$ and store the result in $\mathbf{O}$; (ii) quantization instructions (\texttt{qnt}), which convert extended-precision values in $\mathbf{Y}$ to 8-bit representations; (iii) piecewise-linear (PWL) approximations of non-linear functions (e.g., $\exp$, $\ln$) in the integer domain; (iv) arithmetic operations (add/sub) on post-processing registers; and (v) reduction instructions where either maximum (\texttt{red\_max}) or summation (\texttt{red\_sum}) is applied to all subsequent PPU output results and stored in a given PPR register. To improve throughput, combined activation and quantization instructions (\texttt{qfunc}) are provided. These fuse non-linear transformation and quantization, which is common in DNN workloads (e.g., convolutional layers), where the extended-precision accumulation results must be reduced to 8-bit activations for subsequent layers after the activation function. PPU instructions are typically multi-cycle and support flexible execution through a \emph{blocking} flag. This flag determines whether an instruction must complete before subsequent instructions are issued. When the blocking flag is disabled, instructions can be issued in a pipelined manner, allowing latency hiding and improving overall throughput. Finally, the PPUs also support a multiplication instruction (\texttt{mul}). This is required in normalization operations or when applying per-channel quantization scales.  

\item \textbf{Interrupt}: These instructions enable synchronization and coordination across processing domains, supporting event-driven execution (e.g., signaling the completion of tensor-processing operations to the post-processing units).
\end{enumerate}

The use of the ISA is further illustrated in Section~\ref{s:templates}, which presents code templates for common DNN layers.

\begin{table}[]
\caption{ISA}
\label{t:isa}
\resizebox{\columnwidth}{!}{
\begin{tabular}{@{}llc@{}}
\toprule
\multicolumn{1}{c}{\textbf{Type}}                                                                   & \multicolumn{1}{c}{\textbf{Instruction}} & \textbf{Function}                                                                                                                     \\ \midrule
\multicolumn{1}{c}{\multirow{3}{*}{\textbf{\begin{tabular}[c]{@{}c@{}}Load/\\ Store\end{tabular}}}} &  \texttt{ld} r0, ben, [cp]                              & \begin{tabular}[c]{@{}c@{}} $m_r\gets M[r0]$ \\ben: bank enable, cp: copy flag\end{tabular}                                                                                               \\
\multicolumn{1}{c}{}                                                                                & \texttt{ld\_add} r0,ben, r1, r2, r3, [cp]                        &        $m_r\gets M[r0]$, $r1 \gets r2 + r3$                                                                                                                                \\
      & \texttt{st} r0, r1, ben                              &      $M[r0] \gets r1$                                                                                                                                 \\ 
                      &\texttt{st\_simd} i, j, r0, r1, [tr], [hp] &  $M[r0] \gets \text{pack}(\mathbf{O},i,j,tr,hp)$ $r0\gets r0 + r1$  \\ \midrule
\multicolumn{1}{c}{\textbf{Arith.}}                                                             & \texttt{add} r0,r1,r2,s1,m1,s2,m2                         & $r0 \gets (r1{>>}s1)\&m1 + (r2{>>}s2)\&m2$                \\
                                                                                                    & \texttt{sub} r0,r1,r2,s1,m1,s2,m2         & $r0 \gets (r1{>>}s1)\&m1 -(r2{>>}s2)\&m2$         \\
                                                                                                    & \texttt{addi}  r0,r1,imm                             & $r0\gets r1+imm$    \\
                                                                                                     & \texttt{addic}  r0,r1,imm                             & $r0\gets  (zf==1)~?~r1+imm~:~r1$    \\
                                                                                                    & \texttt{ldi} r0, imm                              & $r0\gets imm$      \\   
                                                                                                     & \texttt{ldid} r0, imm                             & $r0\gets\{imm,imm\}$         \\ \midrule
\multicolumn{1}{c}{\textbf{Branch}}                                                                 & \texttt{b} 'label'                               & unconditional branch to 'label'                                                                                                              \\
                                                                                                    & \texttt{bne} r0,r1,'label                            & branch if $r0\neq r1$                                                        \\
                                                                                                    & \texttt{bnzd} r0, label                            & branch if $r0\neq 0$, $r0\gets r0-1$ \\\midrule
\multicolumn{1}{c}{\textbf{SIMD}}                                                                   & \texttt{mac}                             & $z[i] \gets I_s\cdot I_v[i]+z[i]$ (on next cycle)                                                                                                                                     \\
                                                                                                    & \texttt{pproc}                           & \multicolumn{1}{l}{}                                 $Y \gets Z$, interrupt is generated                                                                                 \\
                                                                                                    & \texttt{qnt} addr, sf                             & $O[addr] \gets \text{Quantize}(Y\cdot2^{sf},8)$                \\
                                                                                                     & \texttt{afunc} 'F', addr                             &  $O[addr] \gets F(O[addr])$                                                                                                                \\
                                                                                                    & \texttt{qfunc} 'F', addr                       & $O[addr] \gets \text{Quantize}(F(Y[addr]),8)$                                                                                                                     \\
                                                                                                     & \texttt{pwl} r0, r1, 'F'             & $r0 \gets F(r1)$                                                      \\
& \texttt{mul} addr, r0             & $O[addr] \gets Y[addr] \cdot r0$                                                      \\
                                                                                                  
                                                                                                    & \texttt{red\_max}                        & initiates max reduction                                                                                                              \\
                                                                                                    & \texttt{red\_sum}                       & initiates sum reduction                                                                                                                  \\ \bottomrule
\multicolumn{1}{c}{\textbf{Interrupt}}                                                               & \texttt{intr\_en} mask                        &  enables/disables interrupts according to mask                                                                                    \\
                   
         & \texttt{intra} i, addr                        &  sets the interrupt routine address for interrupt $i$ \\ \bottomrule
\end{tabular}}
\end{table}

\subsection{Control Core Micro-architecture}

\begin{figure*}[ht!]
    \centering
    \includegraphics[width=1\linewidth]{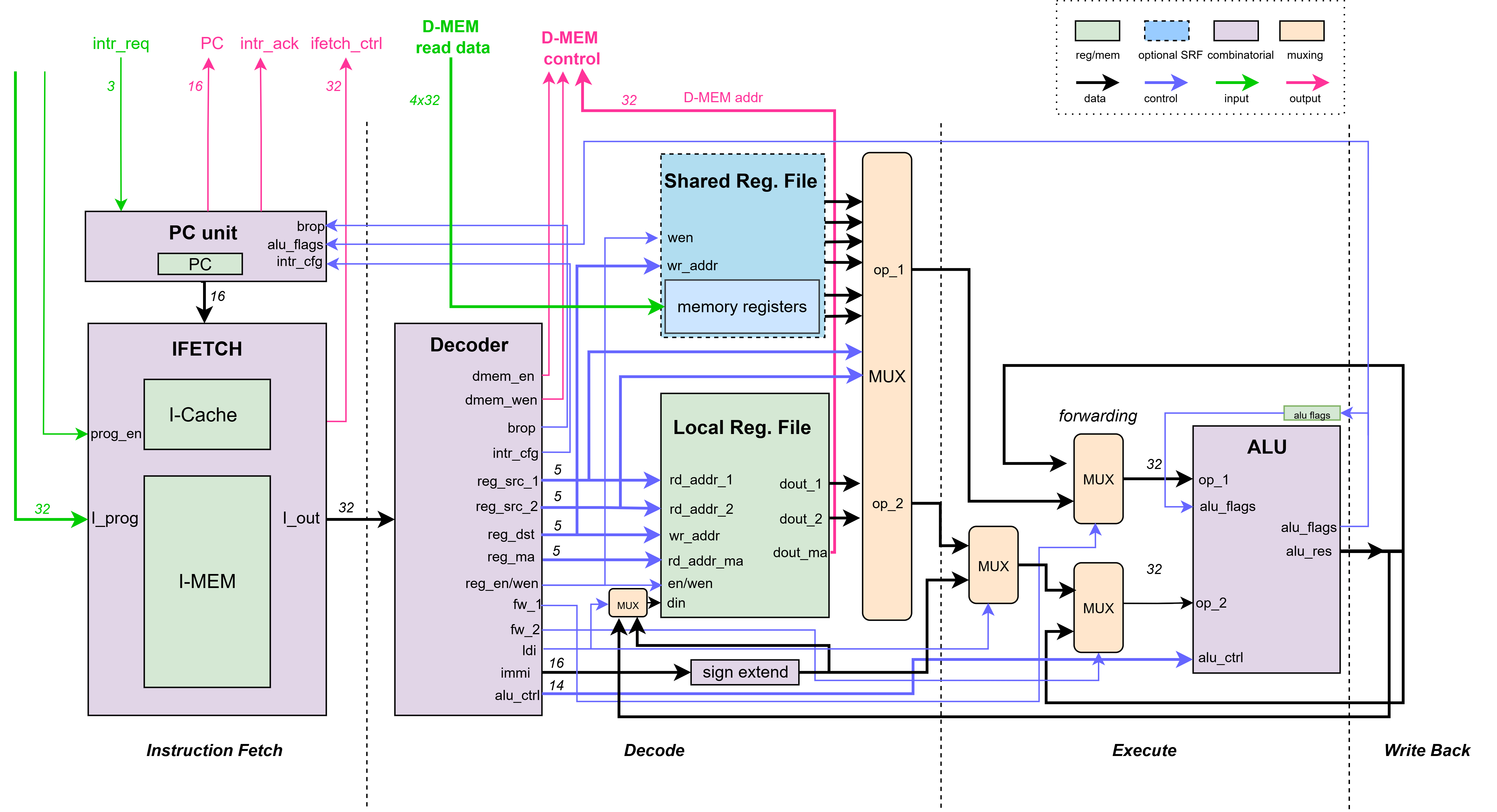}
    \caption{Type-I mCore block diagram. Each mCore features a local register file and an optional shared register file which may be accessible by multiple cores. At the decoding stage,  a decoder block unpacks the instruction fetched from I-MEM during the instruction-fetch stage and generates the necessary control signals. During the execution stage, the ALU unit performs an arithmetic or logical operation on the selected operands and the results are written back to the RF at the final write-back stage.}
    \label{fig:mcore}
\end{figure*}

The proposed processing scheme requires three types of programmable cores (mCores):
\begin{enumerate}
    \item \emph{Type-I (master core):} It supports the full ISA  and it is used to implement the control flow defined by a given DNN layer.
    \item  \emph{Type-II (memory control core)}: It supports a subset of the ISA containing the load/store and arithmetic instructions. It is used in the Data Fetching Domain and can control up to 4 independent memory banks. It does not have an independent program counter (PC); its PC is updated by a type-I core. 
    \item \emph{Type-III (arithmetic core)}: It is used in the pre-processing domain and only supports arithmetic/logical operations.  Its PC is updated by a type-I core. 
\end{enumerate}

The datapath of the mCores is similar to that of RISC cores with four pipeline stages, forwarding, and branch prediction, while their design is customizable: the word length (16/32/64 bits), register file (RF) size (1-32 registers), and number of memory bank interfaces (for type-II cores) can be configured at instantiation time.

 The block diagram of a type-I mCore is shown in Fig.~\ref{fig:mcore}. Type-I mCores produce a program counter (PC) output, which is connected to the PC input of type-II or type-III cores. Type-I and Type-II feature a memory (SRAM) interface that includes read and write enable signals (1 bit per bank of the memory subsystem), an address, and a data field. All mCores feature a local register file, whose depth and word length are configurable,  and an optional shared register file (SRF) interface. An SRF is a collection of registers that can be read by multiple mCores. The SRFs are appended to the register address space of each mCore they are assigned to (at the end of the local register address space); hence, they are visible to the programmer as regular registers. Data memory (D-MEM) outputs are also stored in shared registers and appended to the register address space. The proposed ISA supports 5-bit register addressing; hence, each core can access up to 32 registers (local and shared). Each mCore uses a k-way multiplexer, where k is the size of the SRF, to select the relevant shared register based on the issued instruction. Registers of an SRF can be read simultaneously by multiple cores, but can be updated only by a single core.

The first stage of the pipeline is the instruction fetch (IF) stage, during which instructions are read from the instruction memory (I-MEM). The PC init updates the program counter based on branch evaluation and potential interrupt requests. The I-MEM size in the reference architecture is $0.5$~KB (128 32-bit words). To reduce power consumption, I-MEM is augmented with a small instruction cache (I-cache), implemented as a $64$~B register file (16 registers) using a direct-mapped organization. The I-cache exploits the high temporal locality of DNN workloads, where execution (particularly of the inner loop of a layer routine)  typically spans a small instruction window, to significantly reduce SRAM accesses and thus lower energy consumption. In case of a cache miss, only one additional cycle is required to fetch data from I-MEM.  Since the PC of type-II and type-III cores always follows the PC of a type-I core, cache control signals are only calculated within a type-I core and broadcast to the relevant type-II and type-III cores. Cache hit ratios for various workloads are reported in Section~\ref{s:templates}. A cache-address-enable signal (generated during the cores' programming phase) is also used to optionally mark certain memory addresses as non-cachable, forcing the mCore to skip cache lookups and read from I-MEM directly. This is useful when the number of instructions in the inner loop of a particular DNN layer exceeds the cache size, and we seek to avoid redundant cache miss cycles.

The second stage is the instruction decode (DEC) stage, where the 32-bit instruction fetched from I-MEM is unpacked, and the various control signals are generated. These include RF read/write enable signals and addresses, D-MEM and ALU control signals (such as ALU operation, operand selection, etc.), as well as branching, interrupt, and forwarding control.   

The third stage is the execution stage (EX), where the ALU performs an arithmetic or logical operation on the selected operands. Forwarding logic selects the ALU output as the next ALU operand when a read-after-write dependency exists on a given register. During the execution stage of a \texttt{ld} instruction, a D-MEM read operation at the address pointed by the \textit{dout\_ma} output of the local RF is also initiated. D-MEM output data is written to the corresponding memory registers and hence will be available after 2 cycles. This introduces a read-after-load data hazard on the memory registers, which cannot be avoided. The ALU also generates the zero and negative flags ($zf$ and $nf$), which are used to evaluate branch conditions in the PC unit. These flags are also used in the ALU during conditional-addition (\texttt{addic}) instructions. Finally, the ALU results are written back to the RFs during the write-back stage.


Instructions for the Data Fetching, Pre-processing, and Tensor processing domains are executed in lockstep; thus, one master core (type-I) is used to generate the PC for all three domains, substantially reducing control and decoding complexity on type-II and type-III cores.

\subsection{Baseline Architecture}

In this section, we describe a reference instantiation of the proposed architecture (Fig.~\ref{fig:refernce_arch}). The design includes two type-I mCores: mCore-0 orchestrates the Data Fetching, Pre-processing, and Tensor Processing domains, while mCore-8 controls the Post-processing Domain. Three type-I mCores control the input data streams that feed data from the memory hierarchy to 16 TPA arrays.  Four type-III mCores are used. The post-processing domain consists of 16 PPUs. The system supports two mixed-precision modes with operand precision up to 8 bits.

\subsubsection{Memory Hierarchies}

The baseline architecture employs three memory hierarchies:

\begin{enumerate}[label=\roman*)]
    \item \textbf{OP1MEM}: A $128$\,KB SRAM used to store input tensors and intermediate activations (feature maps). It consists of four 32-bit-wide banks (each bank may be implemented using multiple SRAM macros). Upon a valid read, the output of each bank is written to register $\mathbf{M_1}^{(i)}$, where $i$ is the bank index. OP1MEM supports coarse-grained dynamic precision, storing data in either 8-bit (full precision) or 4-bit (half precision) formats. It is implemented as a pseudo dual-port memory, with one port controlled by mCore-2 and the other by mCore-8.
    
    \item \textbf{OP2MEM}: A $192$\,KB SRAM primarily used for weight storage, but can also be used for storing other types of tensors. It supports fine-grained dynamic precision with operand widths ranging from 3 to 8 bits. Data are stored in a bit-interleaved format across banks. A dedicated hardware unit packs the 32-bit outputs of the $k$ banks into 32 words of $k$ bits each, where $k$ is the selected precision. Further details about dynamic precision are provided in Section~\ref{s:mxp}. OP2MEM outputs are stored in register file $\mathbf{M_2}$ and fed directly to the SIMD units.  OP2MEM is implemented as a pseudo-1R1W memory.
    
    \item \textbf{OP1CACHE}: An $8$\,KB SRAM acting as a software-managed cache for OP1MEM, controlled by mCore-3. While the minimal configuration of the architecture requires only OP1MEM and OP2MEM, this additional memory block demonstrates the extensibility of the data-stream abstraction. In particular, OP1CACHE reduces SRAM accesses during convolution by storing feature-map values that are reused across overlapping computation windows. Its operation is detailed in Section~\ref{s:templates}.
\end{enumerate}

\subsubsection{Data Streams}
Four input and two output DSs are used in the baseline architecture.
The input DSs are: 
\begin{enumerate}[label=\roman*)]
    \item OP1MEM $\rightarrow$ TPAs: Controlled by mCore-1, this DS transfers activation from memory to TPAs. It is processing-enabled; thus, data are processed by type-III mCores-\{4:7\} before being supplied to the TPAs.
    \item OP2MEM $\rightarrow$ TPAs: Controlled by mCore-2, this DS typically transfers weights from memory to TPAs, but can be used for other types of tensors (e.g., for key--value products in attention layers). This DS is not pre-processing enabled.
    \item OP1CACHE $\rightarrow$ TPAs: Controlled by mCore-3, this DS transfers cached OP1MEM operands from memory to TPAs.
    \item OP1MEM $\rightarrow$ PPUs: Controlled by mCore-8, this DS transfers parameters used in post-processing operations (i.e., normalization constants, scaling factors for quantization) from OP1MEM to the PPUs.
\end{enumerate}
The two output DSs transfer output results from the PPUs to either OP1MEM or OP2MEM via a dedicated write-back unit.

\subsubsection{Tensor Processing Arrays (TPAs)}
The architecture comprises 16 tensor processing arrays with 32 processing elements (PEs) each.
Each PE supports dynamic-precision arithmetic operations of up to 8 bits, including multiplication, addition, and fused multiply--accumulate (MAC). To preserve numerical accuracy during accumulation, each PE includes a (16+k)-bit product accumulation register \( \mathbf{z_i} \), and a (16+k)-bit result register \( \mathbf{y_i} \), where $k$ is defined by the user at instantiation time and $i$ is the PE index. Following computation, the final output of each PE is written to an output register file \( \mathbf{O} \) associated with the corresponding SIMD array. The notation $\mathbf{O}_i^j$ denotes the $i$-th output register ($i$ is the index of the PE inside the TPA) of the $j$-th TPA. These register files interface with the Post-processing Domain.

\subsubsection{Registers}
\begin{table*}[ht!]
  \centering
  \caption{Register Files}
  \label{t:register_files}
  \setlength{\tabcolsep}{6pt}
  \renewcommand{\arraystretch}{1.15}
  \begin{tabularx}{0.9\textwidth}{@{}c c c c c c c @{}}
    \toprule
   \textbf{Name} &
   \textbf{Type}&
    \textbf{Num. of regs}&
    \textbf{Bitwidth} &
    \textbf{Accessed by} &
     \textbf{Num. of ports} &
    \textbf{Usage} \\
    \midrule
     $\mathbf{L_1}$ &local       & 8   & 32 & mCore-1                                  & 3                  & local register file \\
    $\mathbf{M_1}$     &memory    & 8   & 32 & mCore-1, mCore-4:7                       & simultaneous access & read data of OP1MEM \\
     $\mathbf{S_1}$   &shared      & 4   & 32 & mCore-1, mCore-4:7                       & simultaneous access & auxiliary regs for OP1MEM  \\    \bottomrule
     $\mathbf{L_2}$   &shared       & 8   & 32 & mCore-2                                   & 3                  &local register file  \\
     $\mathbf{M_2}$   &memory       & 8   & 32 & mCore-2                                   & simultaneous access                   & read data of OP2MEM \\    \bottomrule
     $\mathbf{L_3}$   &local       & 4   & 32 & mCore-3                                   & 3                  & local register file  \\
     $\mathbf{M_3}$   &memory       & 4   & 32 & mCore-3                                   & simultaneous access                   & read data of OP1CAHE \\    \bottomrule
     $\mathbf{L_0}$   &local        & 8   & 32 & mCore-0                                   & 3                  & local register file \\
    $\mathbf{CTRL}$   &shared        & 4   & 16 & mCore-0, TPAs                & simultaneous access   & control and configuration \\    \bottomrule
     $\mathbf{X}$    &shared       & 4   & 32 & mCore-4:7, TPAs                                & simultaneous access                  & pre-processing core outputs \\    \bottomrule
     $\mathbf{L_8}$    &local       & 8   & 32 & mCore-8                                   & 3                 & local register file \\    \bottomrule
     $\mathbf{Z}$   &SIMD        & $N_{\text{PEs}}\!\times\!N_{\text{TPAs}}$ & 20 & TPAs           & simultaneous access & accumulation regs of PEs \\
     $\mathbf{Y}$   &SIMD        & $N_{\text{PEs}}\!\times\!N_{\text{TPAs}}$ & 20 &TPAs, PPUs           & simultaneous access & store the result of PEs in Bext range \\
     $\mathbf{O}$   &SIMD        & $N_{\text{PEs}}\!\times\!N_{\text{TPAs}}$ & 8 &TPAs,PPUs           & simultaneous access & final output  \\    \bottomrule
     $\mathbf{PPR}$ &SIMD   &  $4\times N_{\text{PPU}}$   & 20 & mCore-8                                & simultaneous access & post-processing operations \\
    \bottomrule
  \end{tabularx}
\end{table*}

Table~\ref{t:register_files} lists all registers of the architecture, their size, and functionality. Four types of register files are used in the architecture: local, shared, SIMD, and virtual.
Each mCore-i features is own local register file, $\mathbf{L_i}$. Shared register files correspond to memory output registers or to auxiliary registers accessed by more than one mCore, such as the $\mathbf{CTRL}$ register file, which is updated by mCore-0 and stores various configuration and control parameters (SIMD enable signals, cache configuration, arithmetic precision, padding enable, etc.). SIMD registers store the PE results in the TPAs.

Different SRFs can be grouped together into virtual register files (\textbf{VRFs}). VRFs do not correspond to physical registers; they are groupings of existing physical registers of the system into a continuous address space, used to facilitate reconfigurability and extensibility.  $\mathbf{VRF\_PP}$ is defined as the set of registers corresponding to pre-processing-enabled data streams and contains $\mathbf{M1,M3,S1}$. It is accessible by all pre-processing cores.  If a new DS is to be added to the architecture, a memory subsystem with a type-II mCore and a register file can be instantiated independently. If this DS is pre-processed-enabled, the newly created register file must simply be appended to $\mathbf{VFR\_PP}$ to make it accessible to the pre-processing cores. $\mathbf{VRF\_Iv}$ and $\mathbf{VRF\_Is}$ correspond to the SIMD inputs $\mathbf{I_v}$ and $\mathbf{I_s}$, respectively. In the reference architecture, $\mathbf{VFR\_I_s}$ is formed by grouping the output registers $X_{0:3}$ of the four pre-processing cores, mCore-4:7, while  $\mathbf{VFR\_I_v}$ consists of the OP2MEM  register file $\mathbf{M_2}$.

\subsubsection{Post-Processing Unit (PPU)}
\label{s:ppu}

The reference architecture comprises 16 PPUs, each assigned to a different TPA. Each PPU comprises a dedicated register file, a quantization/scaling unit, a piece-wise linear (PWL) unit, a multiplier, a comparator, and an accumulator (for reduction instructions). The PPUs interface with the $\mathbf{O}$ and $\mathbf{Y}$ TPA register files. They also receive a memory input from the OP1MEM $\rightarrow$ PPU DS, which is controlled by mCore-8, to enable operations directly on memory-resident data. The PPUs support \texttt{ReLU}, \texttt{GeLU}, $\tanh$, \texttt{sigmoid} activation functions. Apart from \texttt{ReLU}, which is implemented in a dedicated hardware unit, all other functions are implemented in the PWL unit. Using the PWL unit, a function $f(x)$ is approximated as a linear function  $f(x)=a_ix+b_i$ within a number of consecutive intervals $n_{\text{int}}$. In the reference architecture, we set $n_{\text{int}}=16$, while $a_i,b_i$ coeffcients are quantized to 12 bits. The PWL receives an 8-bit operand and returns a result in extended 20-bit precision, which can then be passed to the scaling unit for 8-bit quantization. Activation function support can be extended either by adding linear approximations to the PWLs or dedicated hardware for other functions. PPU operation is controlled by a shared register file containing two control registers ($\mathbf{R}$ and $\mathbf{F}$) according to instructions issued by mCore-8.

\subsubsection{Write-back Unit}
\label{s:wb-unit}

When the post-processing domain control core (mCore-8) issues a \texttt{st\_simd} instruction, the write-back unit packs the contents of the 8-bit SIMD output registers $\mathbf{O}$ into 32-bit words before committing them to memory. Three SIMD store modes are supported, selected according to the transpose (\texttt{tr}) and half-precision (\texttt{hp}) flags.

In the default store mode ($hp=0,~tr=0$), the data written to memory bank $i$ is constructed as
\[
o_i \gets \text{concat}\left(\mathbf{O}_j^{4i:4(i+1)}\right),
\]
where $\text{concat}(\cdot)$ concatenates four 8-bit elements into a single 32-bit word, and $j$ denotes the PE register index specified in the \texttt{st\_simd} instruction. The TPA index, which is also specified in the issued instructions, is ignored in this case because concatenation across all TPAs is performed. This organization assumes an output tensor layout compatible with the input layouts defined in the matrix multiplication and convolution templates (Section~\ref{s:templates}), enabling the generated tensor to be directly consumed by subsequent layers without additional data reformatting. In this mode, the write-back unit sustains a bandwidth of four 32-bit words (128 bits) per cycle.

In half-precision mode ($hp=1,~tr=0$), only the upper 4 bits of each SIMD output element are stored. Specifically, for TPA $i$, the upper nibbles of two consecutive output registers, $\mathbf{O}_j^i$ and $\mathbf{O}_{j+1}^i$, are merged into a single 8-bit value. These packed bytes are then concatenated across TPAs to form the final memory words. This mode reduces storage and memory bandwidth requirements for applications that tolerate reduced precision.

The write-back unit additionally supports transpose-store mode ($tr=1$), which writes matrices directly in transposed form during the store operation. In order to store a matrix in transpose format, instead of concatenating output registers across TPAs, registers should be concatenated within the TPAs. In this case, a 32-bit word $o$ will be written to a single bank at a given time, constructed as:
\[
o \gets \text{concat}\left(\mathbf{O}^i_{4j:4(j+1)}\right),
\]
where $i$ and $j$ denote the TPA and PE index, respectively. Hence, in transpose mode, the write-back unit sustains a 32-bit-per-cycle bandwidth.
This functionality is particularly beneficial for attention-based workloads that require matrix transposition. By fusing transpose and store operations, the architecture hides the latency of explicit matrix reordering and avoids additional memory accesses, while incurring only modest hardware overhead in the write-back datapath.

PPU outputs can be written to either OP1MEM (typical operation) or to OP2MEM ($K,Q,V$ matrices for attention). A write-back request router directs write requests to one of these memories, depending on the store instruction's address (the lower part of the address space is assigned to OP1MEM, and the higher part to OP2MEM).  

The write-back unit also features a pooling unit. Maximum and average pooling, typically used in CNNs, is applied along the spatial dimensions of feature maps generated by different PPUs. Thus, pooling is an inter-PPU operation and is implemented within the write-back unit.

\subsection{Program Templates}
\label{s:templates}

\lstdefinelanguage{js}{
  keywords={nop, add, addi, ld, ldcp, st, ldib, b, bnzd, intren, intra, enintr, scl, red_max, red_dis, red_sum, inv, add_ppr, sub_ppr, ldi, scl_ppr, pwl_ppr, pwl_mem, pproc, mac, mac_dis, qfunc, ld_add, addic, subi, sub, ldcp_add, st_add, qnt, add_hpl, add_hph, set_prcs, wait, pwl},
  backgroundcolor=\color{white},   
    commentstyle=\color{codegreen},
    keywordstyle=\color{magenta},
    numberstyle=\tiny\color{codegray},
    stringstyle=\color{codepurple},
    basicstyle=\ttfamily\footnotesize,
    comment=[l]{//},
    morestring=*[d]{"}
}

\lstset{
   language=js,
   extendedchars=true,
   basicstyle=\footnotesize\ttfamily,
   showstringspaces=false,
   showspaces=false,
   tabsize=2,
   breaklines=true,
   showtabs=false,
   numbers=left,                    
    numbersep=5pt,
}
\begin{figure*}[t]
\centering
\begin{tabular}{cccc}
\begin{lstlisting}[language=js]
ldi l2,N/16-1
nop
loop-C: ldi l1 M/32-1 // coulumn loop
loop-R: ldi l0,K-1, // row loop
mac //enable TPAs in next cycle
loop-K: bnzd l0,"loop-K" //inner loop
pproc //send results to PPUs
bnzd l1,"loop-R"
bnzd l2,"loop-C"
\end{lstlisting}&
\begin{lstlisting}[language=js]
ldib l0,K
ldib l2,addrA
ldib l3,0x1
addi l1,l2,0x0
nop
ld_add l1,0xF,l1,l1,l3
nop
nop
add l2,l2,l0
\end{lstlisting}&
\begin{lstlisting}[language=js]
ldib l0,addrB
ldib l3,0x1
ldib l2,K
addi l1,l0,0x0,
nop
ld_add l1,0xF,l1,l1,l3
nop
addi l0,l0,l2
ldib l0,addrB
\end{lstlisting}&
\begin{lstlisting}[language=js]
nop
nop
nop
nop
nop
addi x0,s0
nop
nop
nop
\end{lstlisting}
\end{tabular}
\caption{$(N,K)\times(K,M)$ matrix multiplication routine. From left to right, instructions executed by  mCore-0, mCore1, mCore-2 and mCore-5. mCore-0 implements the control flow (nested loops), enables the TPAs (\texttt{mac} and PPUs (\texttt{pproc}). mCore-1 and mCore-2 generate addresses for OP1MEM (input matrix stored at \textit{addrA}) and OP2MEM (weight matrix stored at \textit{addrB}), respectively. No input prepossessing required, thus mCore-{4:7} execute a trivial routine.}
\label{temp:matmul1}
\end{figure*}

\begin{figure*}[h!]
\centering
\begin{tabular}{llll}
\begin{lstlisting}[language=js]
ldi l3,N/4-1
nop
nop
nop
loop-N: ldi l2, M/4-1
add  l5,l4,l3 //pad logic                     
addic l6,l6,8, 
loop-M: ldi l1, K/32-1
add   l5,l4,l2, 
addic s1,l6,1 //pad reg.
subi  l5,l2,M/4-1 
addic s1,s1,2 //pad reg.
loop-K: ldi l0,C-1
add   l5,l4,1 //cache ctrl.                  
addic s2,l4,1 //cache reg
loop-C: mac
nop //3x3 conv. kernel
nop
nop
nop
nop
nop
nop
nop 
bnzd l0,"loop-C"
pproc
bnzd l1, "loop-K"
bnzd l2, "loop-M"
addi l6,l4,0
bnzd l3, "loop-N"
\end{lstlisting}&
\begin{lstlisting}[language=js]
ldi  l0, bofs
ldib l2, boffs
ldib l3, roffs-boffs
ldib l5, addrI
nop
nop
nop
nop
nop
nop
nop
nop
nop
addi l1,l5,0
nop
nop
ld_add   l1,0xF,l4,l1,l7
add      l2,l1,l2
ldcp_add l2,0xF,l1,l1,l3
ld       l1,0x1
add      l1,l1,l0
ld_add   l2,0x1,s2,m0,l7
ld_add   l1,0x1,l1,l1,l0
ldib     l2,boffs
ldcp_add l1,0x1,l1,l4,l6
nop
nop
nop
add  l5,l5,l2
nop

\end{lstlisting}&
\begin{lstlisting}[language=js]
ldi  l1,boffs
ldi  l3,0
ldib l4,1
ldi  l5,2*boffs
nop
nop
nop
nop
nop
nop
nop
nop
nop
ldcp_add l0,0x1,l0,l3,l6
nop
nop
ld_add   l0,0x3,l2,l0,l6
st_add   l0,0x3,l0,l0,l1
ldcp_add l0,0x1,l0,l0,l5
ld_add   l0,0x1,l6,l6,l6
nop
nop
nop
addi l0,l0, 1
add  l0,l2,l4
sub  l0,l0,l1
sub  l0,l0,l1
nop
add  l3,l3,l1
ldib l3,0

\end{lstlisting}&
\begin{lstlisting}[language=js]
nop
nop
nop
nop
nop
nop
nop
nop
nop
nop
nop
nop
nop
nop
nop
nop
add  x0,s12,s13,3,0x7,3,0x8
addi x0,s12,0
add  x0,s14,s12,1,0xE,1,0x1
add  x0,s4,s13,3,0x7,0,0x8
addi x0,s4,0
add  x0,s4,s0,1,0xE,1,0x1
add  x0,s5,s13,3,0x7,1,0x8
addi x0,s5,0
add  x0,s5,s1,1,0xE,1,0x1
nop
nop
nop
nop
nop
\end{lstlisting}
\end{tabular}
\caption{3x3 convolution routine. From left to right, instructions executed by  mCore-0, mCore-1, mCore-3, mCore-5. mCore-0 implements the control flow, mCore-1 and mCore-3 generate addresses for OP1MEM (input matrix stored at \textit{addrA}) and OP1CACHE (border cbuffer), respectively. mCore-{4:7} concatenate before being supplied to the TPAs. Constants \textit{bofs} and \textit{rofs} are calculated at compile time.}
\label{temp:conv3x3}
\end{figure*}

\begin{figure}[t]
\centering
\resizebox{\columnwidth}{!}{%
\begin{tabular}{cc}
\begin{tabular}{@{}ccc@{}}
\toprule
\multicolumn{3}{c}{\textbf{mCore-1}} \\ \midrule
\textbf{Address} & \textbf{\begin{tabular}[c]{@{}c@{}}Global\\ RF name\end{tabular}} & \textbf{\begin{tabular}[c]{@{}c@{}}Local\\ reference\end{tabular}} \\ \midrule
0x01 & \multirow{4}{*}{L1} & l0 \\
0x02 &                     & l1 \\
$\vdots$ &                 & $\vdots$ \\
0x07 &                     & l7 \\ \midrule
0x08 & \multirow{4}{*}{M1 (shared)} & m0 \\
0x09 &                            & m1 \\
$\vdots$ &                        & $\vdots$ \\
0x0F &                            & m7 \\ \midrule
0x10 & \multirow{4}{*}{S1 (shared)} & s0 \\
0x11 &                             & s1 \\
0x12 &                             & s2 \\
0x13 &                             & s3 \\ \bottomrule
\end{tabular}
&
\hspace{0.5cm}
\begin{tabular}{@{}ccc@{}}
\toprule
\multicolumn{3}{c}{\textbf{mCore-\{4:7\}}} \\ \midrule
\textbf{Address} & \textbf{\begin{tabular}[c]{@{}c@{}}Global\\ RF name\end{tabular}} & \textbf{\begin{tabular}[c]{@{}c@{}}Local\\ reference\end{tabular}} \\ \midrule
0x00 & \multirow{4}{*}{M1 (shared)} & s0 \\
0x01 &                              & s1 \\
$\vdots$ &                          & $\vdots$ \\
0x07 &                              & s7 \\ \midrule
0x08 & \multirow{4}{*}{S1 (shared)} & s8 \\
0x09 &                              & s9 \\
$\vdots$ &                          & $\vdots$ \\
0x0B &                              & s11 \\ \midrule
0x0C & \multirow{4}{*}{M3 (shared)} & s12 \\
0x0D &                              & s13 \\
0x0E &                              & s14 \\
0x0F &                              & s14 \\ \midrule
0x10 &      X (SIMD input)           & x0 \\ 
\bottomrule
\end{tabular}
\end{tabular}%
}
\caption{Register address spaces for mCore-1 and mCore-\{4:7\}.}
\label{fig:regmap}
\end{figure}

A program in the proposed ISA consists of multiple sub-programs, each executed on a different mCore. The Data Fetching, Pre-processing, and Tensor Processing domains operate in a synchronized manner; therefore, the corresponding mCores (mCore-\{1,2\} for data fetching and mCore-\{4:7\} for pre-processing) execute their programs in lockstep. Their program counters are controlled by a master core (mCore-0), which implements the control flow of the current layer. The Post-processing Domain is decoupled and is governed by a separate program executed by a second master core (mCore-8). Programs for each mCore are generated automatically using precompiled templates for common DNN layers and operations. These templates are parameterized according to layer-specific attributes, such as operand memory addresses, layer dimensions, and activation functions. A more general model compiler will be developed in future work. The local register reference used in the following programs for mCore-1 and post-processing mCores-\{4:7\} maps to the global register names according to the tables in Fig.~\ref{fig:regmap}.

\subsubsection{Matrix Multiplication}
The matrix multiplication template is shown in Fig.~\ref{temp:matmul1} and Fig.~\ref{temp:matmul2}. The program computes the product of an $(N,K)$ input matrix $X$ stored at address \textit{addrA} in OP1MEM and a $(K,M)$ weight matrix $\mathbf{W}$stored at address \textit{addrB} in OP2MEM, applies element-wise scaling (8-bit quantization) and ReLU, and stores the result at address \textit{addrC} in OP1MEM. Both inputs and weights use 8-bit precision; thus, each 32-bit word in OP1MEM and OP2MEM stores four values. The assumed memory layout is:

\begin{align*}
    X[i:i+4,j] &\mapsto \text{OP1MEM}\!\left[
        \left\lfloor \frac{i \bmod 16}{4} \right\rfloor
    \right]\!\left[
        \left\lfloor \frac{i}{16} \right\rfloor K + j
    \right] 
\label{eq:dlay0}
\end{align*}
\begin{align*}
    W[i,j:j+4] &\mapsto \text{OP2MEM}\!\left[
        \left\lfloor \frac{j \bmod 32}{8} \right\rfloor
    \right]\!\left[
        \left\lfloor \frac{j}{32} \right\rfloor K + i
    \right],
\end{align*}

\begin{figure}[t]
\begin{lstlisting}[language=js]
ldi l0, addrC 
ldi l0, 0x1
L1:intra 0, "ACTIV" //configure interrupts
wait
ACTIV: intren 0x0 // disable interrupts
qfunc "relu",0x0 // O[0]<- ReLu(quant(Y[0])) 
qfunc "relu",0x1 // O[1]<- ReLu(quant(Y[1])) 
...
qfunc "relu", 0x1F // O[31] <- ReLu(quant(Y[31]))
// store results
st_add 0x0,l0,l0,l1 //mem[l0] <- O[0]; l0 <- l0+l1
st_add 0x1,l0,l0,l1 //mem[l0] <- O[0]; l0 <- l0+l1
...
st_add 0x1,l0,l0,l1
b "L1"
\end{lstlisting}
\caption{Typical post-processing domain program, executed on mCore-8. When accumulation is complete and an interrupt is generated, it applies ReLu to the SIMD array outputs ($\mathbf{Y}$) using fused activation-quantization instructions, and stores the result back in OP1MEM starting at \text{addrC}. Requires 64 cycles to store 32x$N_\text{TPAs}$}
\label{temp:matmul2}
\end{figure}

where the first index corresponds to the memory bank and the second to the row within the bank. The leftmost program (mCore-0) implements the three nested loops required for matrix multiplication using local registers $l0$, $l1$, and $l2$, along with \texttt{bnzd} instructions. Each inner loop iteration calculates a $(N_\text{TPAs}, N_\text{PEs}) = (16,32)$ output block. The \texttt{mac} instruction (Line 5) activates the TPAs in the subsequent cycle, while the \texttt{pproc} instruction (Line 7) forwards the MAC results (stored in the SIMD  registers $\mathbf{Y}$) to the PPU units and generates an interrupt to mCore-8. The inner loop consists of a single instruction (Line 6), during which mCore-1 (OP1MEM) and mCore-2 (OP2MEM) issue memory read operations and compute the next addresses. For OP1MEM,  data from the address pointed to by l1 are fetched and are written to register file $M_1^i$ (where $0\leq i < 4$ is the bank index), which is accessible by the pre-processing mCores. Register l1 is updated to point at the next memory address via a combined load and add instruction (\texttt{ld\_add}). Similarly, mCore-2 issues addresses and fetches weights from OP2MEM. In this matrix multiplication example, no data transformation is required; thus, the pre-processing cores execute a trivial addition, and the results are stored in the register file $X$, which feeds the TPAs. Although pre-processing instructions appear concurrent with data-fetching instructions in the program listing, they are issued with a one-cycle delay to ensure that the referenced registers contain valid data. 

The post-processing control core, mCore-8, executes the program shown in Fig.~\ref{temp:matmul2}. In Line 7, interrupts are enabled, and the interrupt service routine (ISR) is set to start at Line 5. The core then enters an idle state (\texttt{wait}) until an interrupt from the tensor-processing domain—triggered by a \texttt{pproc} instruction—is received.
Upon interrupt, mCore-8 issues 32 \texttt{qfunc} instructions, which are offloaded to the PPUs. Each PPU serially processes the 32 extended-precision PE outputs stored in $\mathbf{Y}$ of its assigned tensor-processing array, applying a ReLU operation followed by 8-bit quantization and saving the result in $\mathbf{O}$.
Finally, the 32 8-bit final outputs are written from register file $\mathbf{O}$ to OP1MEM at the address pointed to by $l0$, which is incremented using \texttt{st\_add}.

\subsubsection{Convolution}
The $3\times3$ convolution template is shown in Fig.~\ref{temp:conv3x3}. The program implements an output-stationary dataflow for the convolution of an $(N,M,C)$ input tensor with a $(3,3,C,K)$ weight tensor, where $C$ and $K$ denote the number of input and output channels, respectively. Output feature maps are computed in a block-wise manner. Each of the 16 TPAs produces one output pixel at a distinct spatial location $(x,y)$. At each cycle, a pixel from the current input tile is broadcast to all 32 PEs of a given array, while each PE receives a different weight corresponding to a distinct output channel.

mCore-0 implements the nested-loop structure of the convolution. The $3\times3$ kernel loop is unrolled (Lines 17--25), with Loop-C iterating over input channels and Loop-K over output channels. Compared to matrix multiplication, additional boundary handling is required; this is implemented using \texttt{addic} instructions to detect feature-map edges (Lines 10, 12) and apply padding. Padding control bits are written to shared register $s1$, corresponding to global padding \textbf{CTRL}. 

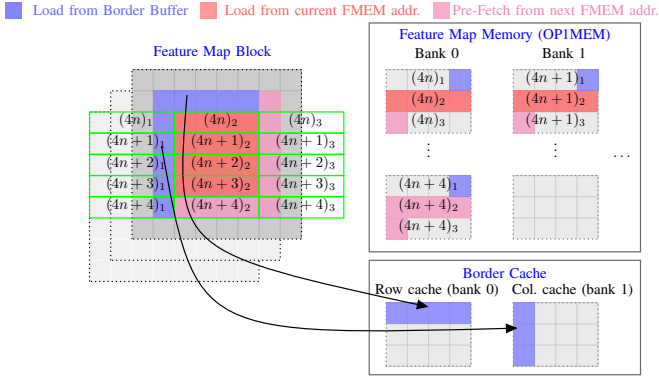
\begin{figure}[t]
    \centering
    \resizebox {1\columnwidth}  {!} {
\input{Figures/conv_mem_layout}}
    \caption{OP1MEM and OP1CACHE organization. Row $(4n+i)_j$ of the feature map block is stored in the $j$-th row of the $i$-th FMEM bank. Pixels that have already been processed are read from the border buffer.}
    \label{f:fmem_layout}
\end{figure}

Data fetching is also more involved. At each step, the memory system must supply both the input pixels of the current block and neighboring pixels from adjacent blocks. For a $3\times3$ kernel, pixels at the borders of the current $4\times4$ block are also required (Fig.~\ref{f:fmem_layout}). Pixels on the top and left borders have already been accessed in previous steps; therefore, to reduce redundant SRAM accesses, a border cache~\cite{sakellariou2024} is employed, consisting of a row and a column buffer. This functionality is implemented via the OP1CACHE data stream, where the column and row buffers are mapped to two distinct banks of OP1CACHE. This buffering scheme avoids re-fetching overlapping input regions, which would otherwise increase memory bandwidth in sliding-window operations. We note that convolution can also be implemented without OP1CACHE, albeit with higher memory traffic.

mCore-1 and mCore-3 generate the required OP1MEM and OP1CACHE addresses, respectively, based on the layout shown in Fig.~\ref{f:fmem_layout}. Feature maps are stored in a row-interleaved format, where the row $(4n+i)$ of a block is mapped to $i$-th FMEM bank. Previously accessed pixels are retrieved from the border buffer, while pixels on the right and bottom edges require additional reads from OP1MEM. In particular, bank 0 must also fetch the bottom row ($(4n+4)$ in the figure), whereas banks 1--3 fetch only the current rows. Consequently, two distinct addresses must be generated per cycle: one for bank 0 and one shared by banks 1--3. To support this, the 32-bit local registers of mCore-1 ($\mathbf{L_1}$) store two 16-bit addresses. The \texttt{ld\_add} instructions fetch data from the address pointed to by $l1$ (Lines 17, 19, 20, 22, 23), while $l1$ is updated according to the block and row offsets (bofs, rofs), which are computed at compile time.  The \texttt{ldcp} instruction preserves the contents of the OP1MEM register in a second set of registers ($\mathbf{M_1}^{(i)},~4\leq i \leq 7$), eliminating redundant memory accesses.

The four pre-processing mCores construct the rows of the $4\times4$ input tile supplied to the TPAs by concatenating pixels from multiple data streams as the $3\times3$ window slides over the input. For example, in the first cycle of the inner loop, the first input row consists of one top-left pixel fetched from OP1CACHE column buffer (stored in $\mathbf{M_3}^{(1)}$ or $s13$ as a local reference) and three rightmost pixels fetched from the row buffer (stored in $\mathbf{M_3}^{(0)}$ or $s12$ as a local reference)  In the subsequent cycle, the input row aligns directly with $s2$, enabling reuse of previously fetched data.

\subsubsection{PE utilization}
\begin{figure}[t]
\input{Figures/util_matmul_conv}
\caption{PE utilization for matrix multiplication (left) with $(N,K,M)$ problem size and convolution (right) on a $64\times64$ input feature map with 64 output channels.}
\label{fig:util1}
\end{figure}
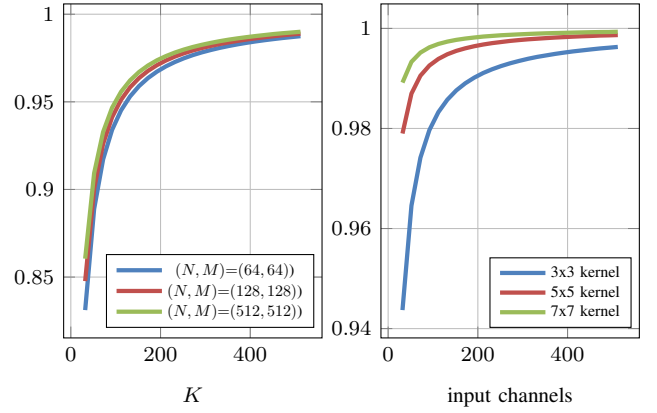

Figure~\ref{fig:util1} shows the PE utilization during the execution of the matrix multiplication and convolution programs for various problem sizes. Utilization increases as the inner loop iterations increase (since this is when MAC instructions are issued). These correspond to the matrix's common dimension or the number of input channels in convolution. For example, a $(64,64)\times (64,64)$ matrix multiplication yields a $90\%$ utilization, while this reaches $98.6\%$ for a $(64,512)\times (512,64)$ multiplication. A convolution on a $64\times64$ input feature map with $64$ output channels ranges from $94.3\%$ to $99.6\%$ for $32 \leq C_{in} \leq 512$  where $C_{in}$ is the number of input channels. These utilization figures account for cache misses as well. In the case of matrix multiplication, the entire program fits on the 16-register I-Cache, hence no misses occur after the cache is filled. In the case of a $3\times3$ convolution, a very high cache hit ratio is also achieved ($>90\%$), since the 9 instructions of the inner loop fit entirely in the cache. 

\subsubsection{Attention layers}
Attention layers compute
\[
\mathrm{Attention}(Q,K,V) = \mathrm{softmax}\!\left(\frac{QK^\top}{\sqrt{d_k}}\right)V,
\]
where $Q$, $K$, and $V$ denote the query, key, and value matrices, obtained via linear projections using weight matrices $W_Q$, $W_K$, and $W_V$, respectively. Since the matrix multiplication routine of \ref{temp:matmul1} multiplies the first operand with the transpose of the second operand, no transpose operation is required to calculate $X=QK^\top$. However, the value matrix $V$ must be stored in transposed format, using \texttt{st\_simd} instructions with $tr=1$.

The softmax operation is computationally expensive and introduces significant latency in hardware accelerators. In the proposed architecture, we approximate softmax using piecewise-linear (PWL) functions in the integer domain and exploit the decoupling of tensor-processing and post-processing domains to hide this latency.

Our approximation is based on the following reformulation:
\begin{align}
    S_i(X) 
    &= \frac{e^{x_i}}{\sum_k e^{x_k}} \nonumber \\
    \ln S_i(X)
    &= \ln\left(\frac{e^{x_i}}{\sum_k e^{x_k}}\right)
     = x_i - \ln\left(\sum_k e^{x_k}\right) \nonumber \\
    &= x_i - x_{\text{max}} - \ln\left(\sum_k e^{x_k - m}\right),
    \label{eq:softmax}
\end{align}
where $x_{\text{max}} = \max_k x_k$ ensures numerical stability. Defining
\[
d = m + \ln\left(\sum_k e^{x_k - m}\right),
\]
we obtain
\[
S_i(X) = e^{x_i - d}.
\]

This formulation requires two exponential evaluations (with outputs in $(0,1]$) and one logarithmic evaluation, all of which can be efficiently approximated using PWL units. Each PWL operation multiplies an $n_i$-bit input by an $n_c$-bit coefficient, producing an $(n_i + n_c)$-bit intermediate result. To prevent dynamic range growth across stages, scaling (i.e., quantization) is applied after each PWL operation.

\begin{algorithm}[t]
\caption{Integer PWL Softmax Approximation}
\label{alg:softmax}
\begin{algorithmic}[1]

\Require Input tensor $X$, PWL parameters
\Ensure Approximated softmax output

\State $X_q \gets \text{Quantize}(X, n_i)$

\State $x_{\max} \gets \max(X_q)$ \Comment{row-wise}
\State $X_q' \gets X_q - x_{\max}$

\State $X_{\exp} \gets \text{PWL\_exp}(X_q')$

\State $s \gets \sum \text{Quantize}(X_{\exp}, n_{i})$
\State $s \gets \text{Quantize}(s, n_{i})$

\State $s_{\ln} \gets \text{PWL\_ln}(s)$

\State $z \gets X_q - (x_{\max} + s_{\ln})$
\State $z \gets \text{clip}(z, \text{minval}, \text{maxval})$

\State $z \gets \text{Quantize}(z, n_i)$

\State $\hat{y} \gets \text{PWL\_exp}(z)$

\State $\hat{y} \gets \text{Quantize}(\hat{y}, n_o)$

\State \Return $\hat{y}$

\end{algorithmic}
\end{algorithm}

Algorithm~\ref{alg:softmax} presents the integer-domain implementation of the proposed softmax approximation, while the implementation of Algorithm~\ref{alg:softmax} in the proposed ISA is shown in Fig.~\ref{fig:attention}. Softmax is applied to the intermediate matrix $X$, of dimensions (N,M), which is obtained using the matrix multiplication routine of Fig.~\ref{temp:matmul1}.  This routine calculates the output matrix in a block-wise manner, where a block is defined as a (16,32) matrix tile. The row-wise maximum $x_{\max}$ is computed on-the-fly using \texttt{red\_max} instructions while the matrix $X$ is generated and stored in a PPR register (Lines 12-16). Once a (16,M) output stripe  of $X$ is available, it is re-fetched to compute the softmax function over two passes. In the first pass (Lines 23-30), the normalization term $s = \sum_k e^{X_k - x_{\max}}$ is calculated.
To efficiently support repeated non-linear transformations on memory-resident data, we introduce the \texttt{pwl\_mem} instruction, which applies a piecewise-linear (PWL) function directly to data fetched from memory and increments the address register. The resulting exponential values are accumulated using \texttt{red\_sum} instructions.
Subsequently, $s_{\ln}$ is computed using a PWL approximation of the logarithm function. Finally, the output is obtained by evaluating $e^z$, where $z = X_q - (x_{\max} + s_{\ln})$,
again using \texttt{pwl\_mem}, followed by quantization to 8 bits (Lines 37-39). Scaling operations required by the algorithm are implemented via \texttt{qnt} instructions.

Thus, for each stripe (16,M), an overhead of $2M$ is paid to apply the softmax, during which the TPAs are idle. Taking Bert-base as an example, where Q $\in \mathbb{Z}^{512 \times 64}, \quad K \in \mathbb{Z}^{512 \times 64}$, the effective PE utilization during attention is $\approx 50\%$. However, taking into account the other matrix multiplications ($Q,K,V$ projections) and MLP in the transformer block, utilization is $\approx 92.5\%$. The online normalization method proposed in ~\cite{onlinenorm} can also be implemented in the proposed ISA, eliminating one of the two input passes required by the softmax algorithm and increasing PE utilization to $67\%$. 

The impact of quantization and PWL approximations of softmax and \texttt{GeLU} on the accuracy of the BERT-base model in the SQuAD dataset is shown in Fig.~\ref{fig:bert_f1}. Moving from 32-bit floating point to an 8-bit integer representation (W8A8) introduces a modest degradation in F1 score of $0.77\%$, while incorporating the approximate softmax and \texttt{GeLU} functions (using PWL approximations of 16 intervals with 12-bit coefficients) results in only a marginal additional loss. These results demonstrate that the proposed approximations preserve accuracy while enabling more efficient hardware implementation.

\begin{figure}
    \centering
    \input{Figures/bert_f1}
    \caption{F1 score of a BERT model on SQUAD under different numerical configurations, including fixed-point (FXP) quantization and the proposed piecewise-linear (PWL) approximations of softmax and \texttt{GeLU}.}
    \label{fig:bert_f1}
\end{figure}
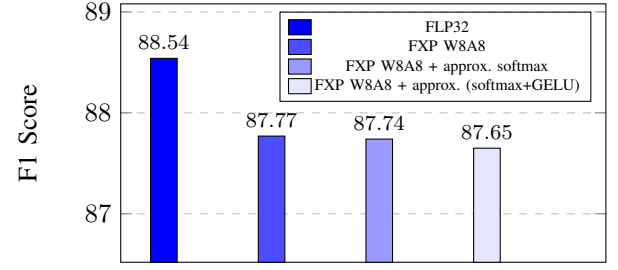
   




\begin{figure}[t]
\begin{lstlisting}[language=js]
ldib l0, addr_tmp // temp store address
ldib l4, addr_tmp
ldib l7, addr_o // output address
ldib l1, 0x1  
ldi  l2, M/32-1 // counter to track row end
ldi  l3, 0x0
intra 0, "SCALE" // set 1st interrupt address
intren 0xF // enabe interrupts
wait
nop
SCALE: intren 0x0 //disable interrupts
red_max r0 // enable max reduction into r0
qnt 0x0, 0x7 // O[0] <- Y[0] / 2^7
qnt 0x1, 0x7 // O[1] <- Y[1] / 2^7
...
qnt 0x1F, 0x7, [blk=1] // blocks execution
st_add 0x0, l0,l0,l1 // mem[l0++] <- O[0]
st_add 0x1, l0,l0,l1 // mem[l0++] <- O[1]
...
st_add 0x1F, l0,l0,l1
bnzd l2, "NOENDROW" // if row end proceed to softmax
red_sum r1 // r1 will store sum(e^((X-max(X)))
add l6, l3, l4
sub_ppr r0, r2, r0, [blk=1] // r0 <- -r0 (r2=0)
addi l5, l3, M
// issue M pwl_mem instructions (full row)
// to calculate row-wise sum(e^((X-max(X)))
SM0:pwl_mem r3,r0,"exp",l6 //r3<-e^(mem[l6++]+r0)
bnzd l5, SMO 
qnt r1,r1,sf1,[blk=1] //quantize sum
pwl r1,r1,"ln",[blk=1] //r2<-ln(sum(e^((X-max(X)))) 
qnt r1,r1,s2f,[blk=1] //qunatize normalization coef.
sub_ppr r1, r2, r1, [blk=1]   // r1 <- -r1 
addi l5, l3,  M 
add l6, l3, l4
// second softmax pass
SM1:pwl_mem r3,r1,"exp",l6 //r3<-e^(mem[l6++]+r1)
st_add r3, l7, l7, l1 // store softmax results
bnzd l5, SM1 
add l4,l3,l0
NOENDROW:
intren 0xF
wait
\end{lstlisting}
\caption{Attention routine. The row-wise maximum of the (N,M) input matrix $X$ is calculated on the fly while $X$ is generated. Then $X$ is re-fetched twice to apply the softmax function according to Algorithm~\ref{alg:softmax} , using \texttt{pwl\_mem} and reduction instructions.}
\label{fig:attention}
\end{figure}

\subsection{Dynamic Precision}
\label{s:mxp}

\begin{figure}
    \centering
    \includegraphics[width=\linewidth]{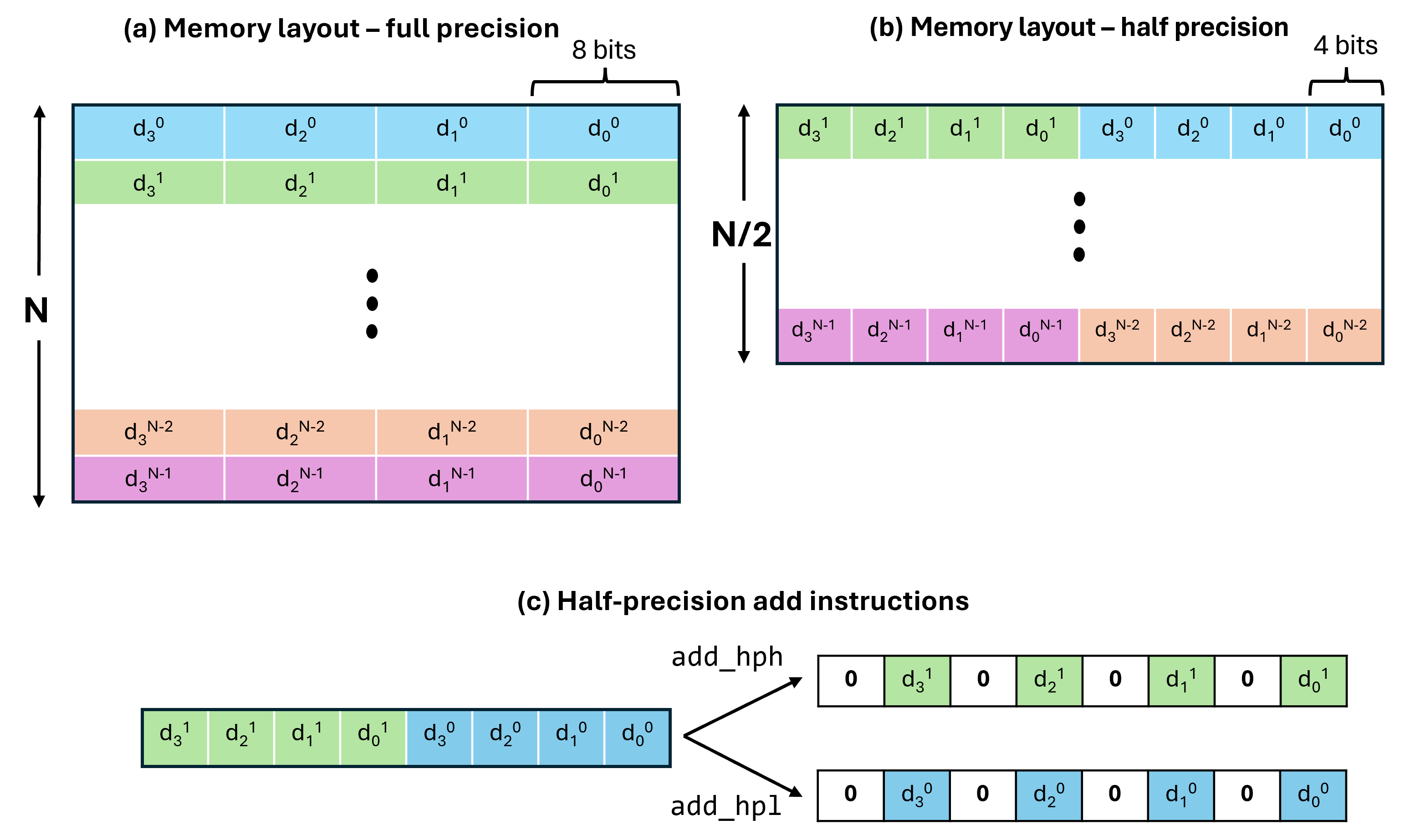}
    \caption{(a) Memory layout during full-precision operation: each 32-bit memory word contains four 8-bit data points. (b) Memory layout during half-precision operation: each 32-bit memory word contains eight 4-bit data points formed by merging consecutive rows in the original memory layout. (c) \texttt{add\_hpl} and \texttt{add\_hph} instructions transform the 16-bit lower or upper part of the 32-bit operand registers into a 32-bit word by interleaving zeros between the four nibbles and then perform regular addition.
    }
    \label{fig:hp}
\end{figure}

\begin{figure*}[t]
\centering
\begin{tabular}{cccc}
\begin{lstlisting}[language=js]
ldi l2,N/16-1
set_prcs 4,4,8 // input,weights,output
loop-C: ldi l1 M/32-1 
loop-R: ldi l0, K/2-1 
mac //enable TPAs in next cycle
loop-K: nop // 2 MACs/loop iteration
bnzd l0,"loop-K" 
pproc //send results to PPUs
bnzd l1,"loop-R"
bnzd l2,"loop-C"
\end{lstlisting}&
\begin{lstlisting}[language=js]
ldib l0,K/2
ldib l2,addrA
ldib l3,0x1
addi l1,l2,0x0
nop
nop //1 load per 2 iter.
ld_add l1,0xF,l1,l1,l3
nop
nop
add l2,l2,l0
\end{lstlisting}&
\begin{lstlisting}[language=js]
ldib l0,addrB
ldib l3,0x1
ldib l2,K
addi l1,l0,0x0,
nop
ld_add l1,0xF,l1,l1,l3
ld_add l1,0xF,l1,l1,l3
nop
addi l0,l0,l2
ldib l0,addrB
\end{lstlisting}&
\begin{lstlisting}[language=js]
nop
nop
nop
nop
nop
add_hpl x0,s0
add_hph x0,s0
nop
nop
nop
\end{lstlisting}
\end{tabular}
\caption{$(N,K)\times(K,M)$ half precision matrix multiplication routine. From left to right, instructions executed by  mCore-0, mCore1, mCore-2 and mCore-4. \texttt{set\_prcs} sets the precision for input (4/8 bits),weights (3-8 bits) and output (4/8 bits). The inner loop iterates K/2 times (instead of $K$) and 2 SIMD MAC instructions are performed at each iteration with only one memory access. Pre-processing cores issue \texttt{add\_hpl} and \texttt{add\_hph} instructions to supply the lower and upper part of the 32-bit input fetched from memory to the TPAs  at alternating cycles, respectively. }
\label{temp:matmul_hp}
\end{figure*}

Dynamic-precision (DP) quantization exploits the fact that the minimum numerical precision required to preserve accuracy varies across neural network layers. By assigning different bit-widths to weights and activations on a per-layer basis, DP quantization reduces hardware cost while maintaining acceptable accuracy. The proposed architecture supports two forms of DP: (a) \emph{coarse-grained} and (b) \emph{fine-grained}.

Coarse-grained DP targets input and intermediate activations and supports either 8-bit or 4-bit precision. OP1MEM and OP1CACHE operate in this mode. Half-precision operation is exposed at the ISA level through the \texttt{add\_hpl} and \texttt{add\_hph} instructions. In 4-bit mode, each 32-bit memory word stores eight 4-bit values, allowing a single memory access to supply data for two consecutive SIMD cycles. The pre-processing cores unpack these values by issuing successive \texttt{add\_hpl} and \texttt{add\_hph} instructions, operating on the lower and upper halves of the register, respectively (Fig.~\ref{fig:hp}).  This scheme halves memory reads while preserving the original memory organization and requiring no additional hardware.

Fine-grained DP is applied to read-only network parameters (weights) and supports precisions from 3 to 8 bits. In the reference architecture, OP2MEM implements this mode by storing weights in a bit-interleaved format across memory macros. A memory access activates a number of macros proportional to the selected precision, while dedicated hardware packs the retrieved bits into $k$-bit words, where $k$ is the bit-precision. Since weights are read-only, this approach provides greater flexibility without incurring write-back overhead.

Operand precisions are configured through dedicated \texttt{set\_prcs} instructions that update the precision-control registers of $\mathbf{CTRL}$. Both DP schemes result in memory power consumption that scales approximately linearly with bit precision. The corresponding matrix-multiplication templates are shown in Fig.~\ref{temp:matmul_hp}.  The inner loop iterates $K/2$ times (instead of $K$) and 2 SIMD MAC instructions are performed at each iteration with only one memory access. 

\subsection{Layer Fusion}
Depthwise (DW) separable convolution, widely used in lightweight models such as MobileNet~\cite{mobilenet}, significantly reduces computational complexity and parameter count compared to conventional convolution. However, DW convolution often leads to low processing-element utilization in DNN accelerators~\cite{dnnaccel6} because each filter operates on a single input channel, greatly limiting opportunities for data reuse. As a result, the computation-to-memory-access ratio is substantially lower than in standard or pointwise (PW) convolutions, making performance increasingly constrained by memory bandwidth rather than arithmetic throughput.

To improve utilization for MobileNet-like workloads, we exploit the observation that each DW convolution is immediately followed by a PW convolution with $1\times1$ kernels. Instead of executing the two layers sequentially and storing the intermediate DW outputs in OP1MEM, we fuse them into a single processing pipeline. Specifically, the PEs within each TPA can be partitioned into two groups: DW PEs and PW PEs. After a $4\times4$ input tile is processed by the DW PEs, the resulting feature map block is buffered in a small local memory and streamed directly to the PW PEs. Meanwhile, the DW PEs proceed to the next input tile, enabling concurrent execution. Since PW convolution performs cross-channel accumulation and dominates the computational workload, a larger fraction of the available PEs is assigned to the PW stage. The resulting utilization depends on both the number of PW output channels and the allocation of PEs between the DW and PW groups. By overlapping DW and PW execution, the proposed scheme substantially improves PE utilization and mitigates the memory-bound nature of depthwise separable convolutions.

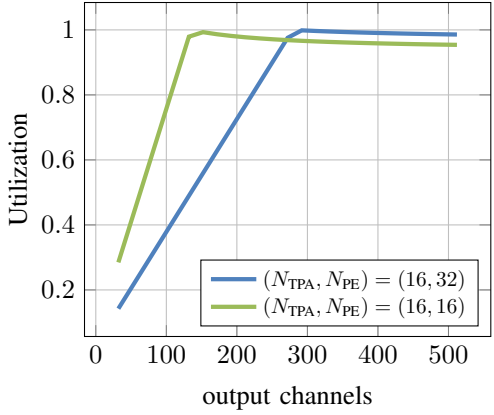
\begin{figure}
    \centering
    \input{Figures/util_dw_pw}
    \caption{Fused DW-PW utilization as a function of output channels}
    \label{fig:util_pw_dw}
\end{figure}

To support layer fusion, we introduce an additional input DS from the DW output buffer (OP1FUSED) to the TPAs, controlled by a new type-I mCore. This DS is not pre-processed enabled and thus the destination registers of OP1FUSED (which is organized in four banks as OP1MEM), are directly  appended to $\mathbf{VRF\_Is}$. Inside each TPA, we set the number of DW PEs $g_{DW}$ to 1 and the number of DW PEs $g_{PW}= 32 -g_{DW}=31$. PEs in the PW group always receive data from the OP1MEM original DS, while the DW can conditionally receive their data from the newly added OP1FUSED stream during the depthwise mode. Thus, each DW PE uses a multiplexer to select between the two. The \texttt{mac} instruction will now also specify an enable flag for each of the two PE groups. The number of cycles required for the completion of fused DW-PW convolution for a $4\times4$ input tile with $C_{in}$ input channels and $K$ output channels is given by:
\[C_{in}\cdot\text{max}(\frac{9}{ N_{TPA}\cdot g_{DW}},\frac{K}{N_{TPA}\cdot g_{PW}}),\]
assuming a $3\times 3$ DW kernel.

The layer fusion scheme demonstrates the flexibility of the proposed architecture, enabling new features to be added with minimal control overhead and design effort.

\section{RNS Instantiation}

This section presents a Residue Number System (RNS) instantiation of the proposed architecture that improves the performance of the baseline fixed-point (FXP) instantiation. The ISA is designed so that control mCores execute the same program regardless of the underlying data representation, with FXP-to-RNS conversion taking place at the hardware level (not exposed to the programmer).

\subsection{RNS Basics}

The Residue Number System (RNS) is an alternative numerical representation in which an integer is encoded by its residues with respect to a set of pairwise coprime moduli $\{m_0,m_1,\ldots,m_{n-1}\}$, referred to as the \textit{base}. An integer $X$ is represented as
\[
X \mapsto (x_0,x_1,\ldots,x_{n-1}), \qquad x_i = \langle X \rangle_{m_i},
\]
where $\langle \cdot \rangle_{m_i}$ denotes the modulo-$m_i$ operation. The dynamic range of the representation is given by the product of the base moduli. Due to the mathematical properties of the modulo operator, addition and multiplication are performed independently in each residue channel; thus, RNS arithmetic eliminates long carry-propagation paths, enabling higher operating frequencies and improved energy efficiency~\cite{rns1}. Other operations, such as division and comparison, have non-trivial RNS implementations and typically introduce hardware overhead. However, in typical DNN workloads, the number of these operations is significantly lower (2-3 orders of magnitude) than the number of multiply-add operations. Consequently, RNS-based DNN accelerators have demonstrated significant advantages over conventional fixed-point (FXP) designs~\cite{rns1,rns2,rns3,rns4,rns5,rns6,rns8}.

\subsection{Dynamic Precision in RNS}
\label{s:rns-mxp}

Although RNS has traditionally been associated with large word-length computations, recent work has demonstrated that RNS-based accelerators can also outperform FXP counterparts in a low-precision ($3$--$8$ bit) DP quantization regime, consistently achieving superior accuracy--hardware cost tradeoffs across a range of DNN models ~\cite{iscas2025sakel}.

IN DP quantization, the objective is to determine layer-wise precision configurations that minimize the total inference cost $C$ under a target accuracy degradation constraint $\delta L_{\text{target}}$, or equivalently, maximize accuracy under a cost budget. Assuming inter-layer independence~\cite{mxp1}, the optimization problem is expressed as
\[
C = \sum_{i=1}^{L} C_i,
\qquad
\delta L = \sum_{i=1}^{L} \delta L_i < \delta L_{\text{target}},
\]
where $L$ is the number of layers, and $C_i$ and $\delta L_i$ denote the hardware cost and accuracy degradation associated with quantizing layer $i$, respectively.
Various algorithms can be employed to solve this optimization problem and perform the initial quantization in a fixed-point format. Here we employ the OBC~\cite{obc} method for quantization and solve the constraint optimization problem using dynamic programming.

DP quantization can then be extended to the RNS domain by dynamically selecting subsets of a primary RNS base for performing arithmetic operations at each layer. Let $\mathcal{B}_P$ denote the primary RNS base, chosen to cover the maximum required dynamic range. Different fixed-point quantization configurations are then mapped to subsets $\mathcal{B}_i \subseteq \mathcal{B}_P$ of equivalent dynamic range (determined by the product of their constituent moduli).  Quantization formats are denoted using the \emph{WxAy} notation, where $x$ and $y$ represent the fixed-point bit-widths of weights and activations, respectively.

We assume a primary base
\[
\mathcal{B}_P = \{5,7,9,31,32\},
\]
which efficiently covers the W8A8 (INT8) range.
The quantization mapping is defined as follows:
\begin{itemize}
    \item W8A8, W7A8 $\mapsto \mathbf{\mathcal{B}_0 = \{5,7,9,31,32\}} \quad (\approx 18\text{ b})$
    \item W6A8 $\mapsto \mathbf{\mathcal{B}_1 = \{7,9,31,32\}} \quad (\approx 16\text{ b})$
    \item W5A8, W4A8 $\mapsto \mathbf{\mathcal{B}_2 = \{5,7,31,32\}} \quad (\approx 15\text{ b})$
    \item W3A8, W4A7 $\mapsto \mathbf{\mathcal{B}_3 = \{5,7,9,32\}} \quad (\approx 13\text{ b})$
    \item W4A4, W3A4 $\mapsto \mathbf{\mathcal{B}_4 = \{5,7,32\}} \quad (\approx 10\text{ b})$
\end{itemize}

We consider quantization configurations with $x \in \{3,4,\ldots,8\}$ and $y \in \{4,8\}$, excluding intermediate activation precisions  to avoid excessive implementation complexity. For each configuration, the selected base subset provides a dynamic range
\[
R \geq 2^{x+y+2},
\]
allowing the accumulation of partial products with negligible overflow error \cite{iscas2025sakel}.

To apply the DP quantization methodology in the RNS setting, we use an RNS-aware hardware cost function that estimates the total energy cost of model execution on the proposed architecture (Fig.~\ref{fig:refernce_arch}), based on power consumption estimates for the various components. The energy cost per inference is:
\begin{equation}
    C = \sum_{l=1}^{L}\left[{\sum_{j=1}^{\operatorname{size}(\mathcal{B}_0)}}\!\!\!P_jz_j^{(l)} {+} \alpha_m^{(l)}P_{\text{mem}} {+} \alpha^{(l)}_aP_{\text{ppu}}\right]\!\text{O}_l\frac{T_{\text{clk}}}{N_{\text{PE}}}
\label{eq:metric}    
\end{equation}
where $P_j$, $P_\text{mem}$, and $P_\text{ppu}$ correspond to the power consumption of the $j$-th residue channel of the TPAs, the memory system,  and the PPUs, respectively, at clock period $T_{\text{clk}}$.  O$_l$  denotes the number of operations of layer~$l$, $N_{\text{PE}}$ is the total number of PEs in the accelerator, $z_j^{(l)}$ are the binary variables evaluated by the optimization procedure, that denote whether channel $j$ is used at layer~$l$, $ \alpha^{(l)}_a$ determines the contribution of the power consumption of the PPUs according to their utilization ratio and the $z_j^{(l)}$ values, and $\alpha_m^{(l)}$ determines the contribution of the memory and depends on the $x$,$y$ values of the WxAy configurations.

\subsection{RNS Operations}

\subsubsection{TPAs}
RNS TPAs consist of parallel and independent multiply-accumulate residue channels. For each layer, only the subset of channels required by the selected precision is activated, while inactive channels are clock-gated to eliminate dynamic power consumption. Standard low-cost RNS arithmetic techniques are employed: end-around-carry adders for moduli of the form $2^k{-}1$ and diminished-1 arithmetic for moduli of the form $2^k{+}1$~\cite{dim1mult}. A methodology for systematically selecting optimal RNS bases under DP quantization is presented in Section~\ref{s:base_sel}.

\subsubsection{Conversion}
Operands are stored in memory in fixed-point format and converted to RNS on the fly. Consequently, a conversion stage is inserted between the SIMD input registers and the TPAs in Fig.~\ref{fig:refernce_arch}. Although this stage introduces a one-cycle delay to all input data streams, it does not affect throughput. The use of low-cost moduli significantly reduces conversion complexity. The corresponding power and area overheads are reported in Table~\ref{tab:power_breakdown}.

\subsubsection{PPUs}
RNS PPUs support the instruction set described in Section~\ref{s:isa}. Certain operations, such as \texttt{afunc} and \texttt{qnt}, require more complex RNS circuitry and therefore incur higher latency. However, this overhead is effectively hidden by the decoupled tensor-processing and post-processing domains and by the high MAC-to-PPU operation ratio.

The core PPU component is a dynamic base-extension unit, used for power-of-two scaling, sign detection, overflow handling, and interval selection in piecewise-linear activation functions~\cite{iscas2025sakel}. This approach avoids dedicated circuitry for each base subset and reduces power consumption through clock-gating of inactive registers. As a result, the dynamic power of the unit scales approximately linearly with the number of active bits.

An additional \texttt{bext\_acc} instruction is introduced to support higher dynamic-range accumulation by extending the bitwdith of the power-of-two channel (32) to 8 bits (256). This instruction performs accumulation with overflow checking and saturation, mitigating overflow-induced accuracy degradation~\cite{iscas2025sakel}. Finally, the post-processing instruction-blocking/non-blocking mechanism enables mCore-8 to execute the same program regardless of instruction latency variations.

\subsection{Optimal Base selection}
\label{s:base_sel}

\begin{figure*}[t]
\centering
\subfloat[ ResNet-18][ ResNet-18]{
    \resizebox {0.31\textwidth}  {!} {
        \input{Figures/base_comp_rn18}
    }
}
\hfill
\subfloat[ResNet-50][ResNet-50]{
    \resizebox {0.31\textwidth}  {!} {
        \input{Figures/base_comp_rn50}
    }
}
\hfill
\subfloat[Yolov5-m][Yolov5-m]{
    \resizebox {0.31\textwidth}  {!} {
        \input{Figures/base_comp_yolo}
    }
}
\caption{Normalized energy cost vs model loss constraint for the identified $N_2$ \emph{optimal} bases, for (a) ResNet-18 and (b) ResNet-50 models.}
\label{fig:basecomp_mxp}
\end{figure*}
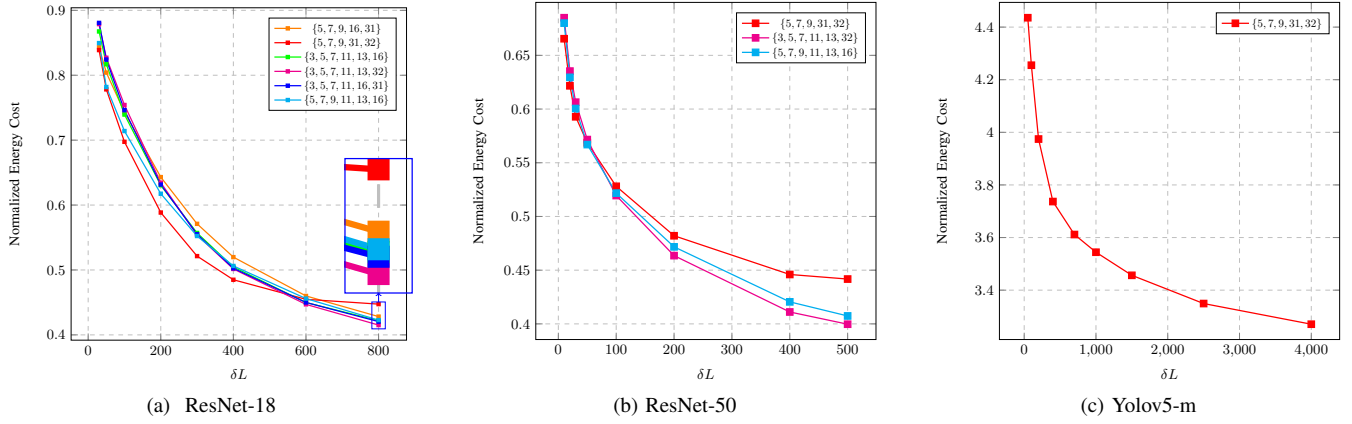

Existing RNS-based DNN accelerators~\cite{rns1,rns2,rns3,rns4,rns5,rns6,rns8} typically select RNS bases heuristically, favoring low-cost moduli such as $2^k$ and $2^{k\pm1}$ that simplify hardware implementation. In contrast, we propose a systematic methodology for identifying optimal RNS bases (or base families) for dynamic-precision DNN workloads. Using the hardware-aware mixed-precision quantization framework of Section~\ref{s:rns-mxp}, we classify RNS bases as either \emph{sub-optimal}, i.e., consistently outperformed by another base, or \emph{optimal}. Starting from the modulus pool
\[
P=\{3,4,5,7,8,9,11,13,15,16,17,19,23,31,32,33,64\},
\]
we generate all valid co-prime subsets with dynamic range
\[
2^9 \leq R \leq 2^{20},
\]
while requiring at least one power-of-two modulus to reduce hardware complexity. This results in $N_0{=}2123$ candidate bases.

For each modulus $m\in P$, the corresponding MAC unit is synthesized to obtain its hardware cost $h(m)$ (power under fixed voltage and timing constraints). Given a primary base $B$, each quantization configuration WxAy is mapped to the lowest-cost subset satisfying the required dynamic range:
\[
F_B(x,y)=
\argmin_{
\substack{
b_i\in B\\
R(b_i)\geq 2^{x+y+2}
}}
a\sum_{m\in b_i} h(m)+b\,h_{\text{aso}}(b_i),
\]
where $h_{\text{aso}}(b_i)$ estimates the cost of auxiliary RNS units (activation, scaling, overflow handling), and $a,b$ weight the relative contributions of MAC and post-processing hardware. Since $h_{\text{aso}}(b_i)$ scales approximately linearly with the total bit-width of $b_i$, bases with unnecessarily large dynamic ranges are penalized. To reduce the search space, we first eliminate strictly inferior bases, i.e., bases for which another candidate yields lower hardware cost for all WxAy mappings. This filtering reduces the set to $N_1=27$ candidates. The DP optimization algorithm is then executed for each base to obtain loss--cost pairs $(\delta l_i,c_i)$ across different target accuracy degradations. A base is classified as \emph{sub-optimal} if another base achieves lower cost for all $\delta l_i$ values; otherwise, it is considered \emph{optimal}.

Using ResNet-18 data, the procedure identifies $N_2=6$ optimal bases (Fig.~\ref{fig:basecomp_mxp}-(a)), assuming a $0.65$~V supply and $1$~ns timing constraint. Two optimal five-channel bases are obtained, consisting of $\{5,7,9,31\}$ combined with either $16$ or $32$, together with four six-channel bases. Three of the latter are formed by combining $\{3,5,7,11\}$ with two moduli selected from $\{13,16,31,32\}$, while $\{5,7,8,11,13,16\}$ is also identified as optimal. Similar trends are observed for ResNet-50. The optimal base selection depends on the synthesized hardware costs of the residue channels and therefore varies with timing constraints. For example, under a tighter $700$~ps constraint, modulus $17$ replaces $13$. Although six-channel bases provide better efficiency at the low-cost end for ResNet models, the five-channel base $\mathcal{B}=\{5,7,9,31,32\}$ dominates across all loss constraints for YOLO-v5 while also offering lower area overhead. Overall, this analysis identifies a compact and robust RNS base suitable for a broad range of DP DNN workloads.

\section{Results}
In this section, we report the implementation results (power and area) of the RNS instantiation after a full place-and-route run. We compare this instantiation with the conventional FXP counterpart under the dynamic precision optimization framework for varying accuracy constraints. Finally, we report qualitative and end-to-end quantitative comparisons with state-of-the-art fixed-function and RISC-V-based accelerators.

\subsection{Implementation Results}

\begin{figure}
    \centering
    \includegraphics[width=1\linewidth]{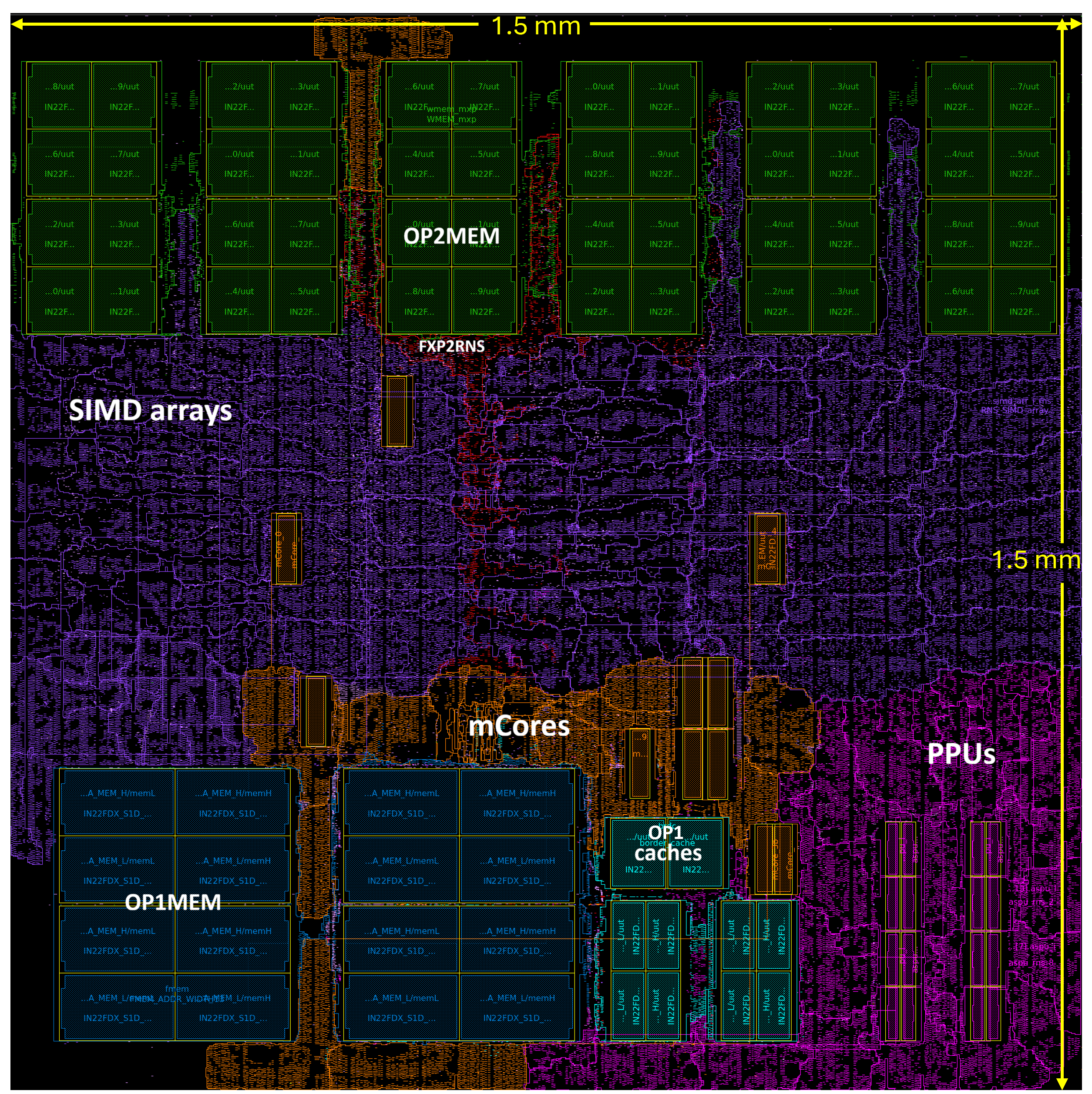}
    \caption{Placed and routed design with annotated hierarchy. }
    \label{fig:floorplan}
\end{figure}

We implement the proposed RNS-based architecture on a 22~nm FDX GlobalFoundries technology using Synopsys IC Compiler II. The placed-and-routed view of the design with annotated hierarchy is shown in Fig~\ref{fig:floorplan}. The core area is $2.25$~ mm$^2$ ($1.5$~mm $\times$ $1.5$~mm). Timing closure for nominal operating conditions of $0.8$~V is verified at $800$~MHz.  

\begin{table}[h]
\centering
\caption{Power breakdown for \texttt{matmul} $(64,512)\times(512,64)$ at 0.65\,V, 500\,MHz (98\% utilization).}
\label{tab:power_breakdown}
\scriptsize
\setlength{\tabcolsep}{3pt}
\begin{tabular}{lcccccc}
\toprule
& \multicolumn{2}{c}{\textbf{W8A8}} 
& \multicolumn{2}{c}{\textbf{W4A7}} 
& \multicolumn{2}{c}{\textbf{W3A4}} \\
\cmidrule(lr){2-3} \cmidrule(lr){4-5} \cmidrule(lr){6-7}
\textbf{Unit} & mW & \% & mW & \% & mW & \% \\
\midrule

{16}$\times${TPAs}  & 50.40 & 51.6 & 24.16 & {39.59} & 19.36 & 40.51 \\

{16}$\times${PPU}  & 3.36 & 3.44 & 2.56 & 4.19 & 1.60 & 3.34 \\
{FXP-to-RNS}         & 3.86 & 3.95 & 2.56 & 4.19 & 2.05 & 4.28 \\

\textbf{Total SIMD}      & 57.62 & \textbf{59.06} & 29.28 & \textbf{47.98} & 23.01 & \textbf{48.14} \\

\midrule

mcore-0         & 0.62 & 0.63 & 0.62 & 1.01 & 0.62 & 1.29 \\
mcore-1         & 0.54 & 0.55 & 0.54 & 0.88 & 0.65 & 1.36 \\
mcore-2         & 0.52 & 0.54 & 0.52 & 0.85 & 0.52 & 1.08 \\
mcore-4         & 0.35 & 0.36 & 0.35 & 0.57 & 0.38 & 0.79 \\
mcore-5         & 0.35 & 0.36 & 0.35 & 0.57 & 0.38 & 0.79 \\
mcore-6         & 0.35 & 0.36 & 0.35 & 0.57 & 0.38 & 0.79 \\
mcore-7         & 0.35 & 0.36 & 0.35 & 0.57 & 0.38 & 0.79 \\
mcore-8         & 0.93 & 0.95 & 0.93 & 1.51 & 0.93 & 1.93 \\

\textbf{Total control}   & 4.02 & \textbf{4.11} & 4.02 & \textbf{6.58} & 4.24 & \textbf{8.86} \\

\midrule
OP1MEM            & 7.45  & 7.63  & 7.46 & 12.27 & 4.35 & 9.10 \\
OP2MEM            & 13.59 & 13.39 & 7.66 & 12.55 & 6.20 & 12.96 \\
OP1(CACHE+FUSED)          & 0.1  & 0.1  & 0.16 & 0  & 0 & 0.2 \\

\textbf{Total Mem.}       & 20.81 & \textbf{21.33} & 16.62 & \textbf{24.94} & 11.55 & \textbf{22.27} \\

\midrule

\textbf{Clock Network}   & 15.10 & \textbf{15.47} & 12.50 & \textbf{20.45} & 9.90 & \textbf{20.70} \\

\midrule
\textbf{Total}           & \textbf{97.5} & & \textbf{61.01} &  & \textbf{47.88} &\\

\midrule
\textbf{Power Eff. (TOPS/W)}      
& \multicolumn{2}{c}{\textbf{5.14}}
& \multicolumn{2}{c}{\textbf{8.22}}
& \multicolumn{2}{c}{\textbf{10.47}} \\
\bottomrule
\end{tabular}
\end{table}

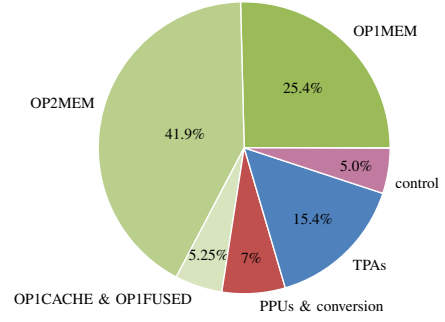
\begin{figure}
    \centering
    \resizebox{!}{0.22\linewidth}{
    \input{Figures/area}}
    \caption{Area breakdown}
    \label{fig:area_breakdown}
\end{figure}

The area breakdown is shown in Fig.~\ref{fig:area_breakdown}.  The memory hierarchy accounts for the majority of the system's area, with OP2MEM contributing the largest share (41.9\%), followed by OP1MEM (25.4\%). In comparison, TPAs account for 15.4\%, while the combined PPU and conversion logic account for 7\%. Control logic accounts for only 5\% of the total area. 

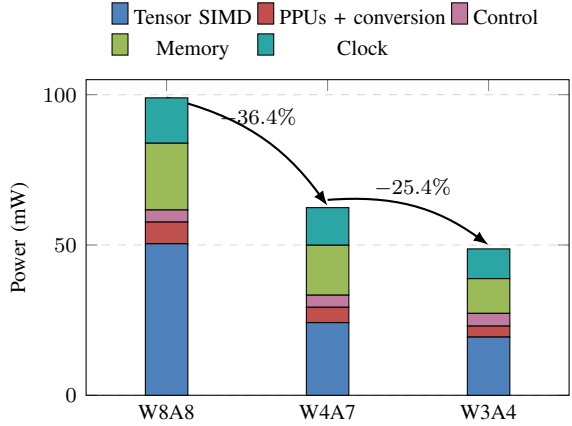
\begin{figure}
    \centering
    \input{Figures/power}
    \caption{Main power contributors under W8A8, W4A7, and W3A4 quantization.}
\label{fig:power_breakdown}
\end{figure}

We report power analysis results during low-power operation, where the system operates at $0.65$~V and $500$~MHz. We use an indicative $(64,512)\times(512,64)$ matrix-multiplication workload and collect switching activity statistics across W8A8, W4A7, and W3A quantization configurations (that use $B_0$, $B_3$, and $B_4$ RNS bases respectively) to enable accurate power analysis. The power breakdown (Table~\ref{tab:power_breakdown} and Fig.~\ref{fig:power_breakdown}) indicates that TPAs are the dominant contributor across all quantization modes, accounting for 51.6\%, 39.6\%, and 40.51\% of the total power in W8A8, W4A7, and W3A4, respectively. In contrast, the control overhead remains relatively small, particularly for W8A8 (4.1\%), but becomes more significant at lower precisions, reaching 8.9\% in W3A4. This trend arises because the mCore power consumption is independent of the quantization configuration, whereas the SIMD datapath and memory power consumption scales with precision. As the precision decreases from W8A8 to W3A4, the total power consumption is reduced substantially from 97.5\,mW to 47.9\,mW, while the power efficiency improves from 5.14 to 10.47\,TOPS/W. Memory accounts for approximately 21.3--24.9\% of the total power, and the clock network contributes around 15--21\% across all configurations. OP1CACHE is not used during matrix multiplication, and thus the corresponding core is clock-gated in this workload. 
These results confirm that the principal design goal of introducing minimal control overhead through lightweight control mCores, while maintaining a high level of programmability, is achieved.  

\subsection{RNS vs FXP Comparison}

\definecolor{bblue}{HTML}{4F81BD}
 \definecolor{rred}{HTML}{C0504D}
 \definecolor{ggreen}{HTML}{9BBB59}
\begin{figure*}[t!]
\centering
\hfill
\subfloat[][]{
    \resizebox {0.48\textwidth}  {!} {
        \input{Figures/perf_rn50}
    }
}
\hfill
\subfloat[][]{
    \resizebox {0.48\textwidth}  {!} {
        \input{Figures/perf_yolo5m}
    }
}
\hfill
\subfloat[][]{
    \resizebox {0.48\textwidth}  {!} {
        \input{Figures/perf_bert}
    }
}
\subfloat[][]{
    \resizebox {0.48\textwidth}  {!} {
        \input{Figures/perf_vit16}
    }
}
\caption[MXP Results]{Normalized (with respect to a uniform W8A8 ($\mathcal{B}_0$) RNS baseline) energy cost as a function of the achieved accuracy. Black dashed lines show the baseline FLP32 performance, while \textcolor{magenta}{pink} dashed line corresponds to a 1\% accuracy, F1 or mAP drop. The RNS mixed-precision setting (\textcolor{ggreen}{green}) can either match or outperform all uniform quantization configurations (\textcolor{rred}{red}) as well as the FXP mixed-precision setting (\textcolor{bblue}{blue}) for all accuracy drop constraints.}
\label{fig:performance}
\end{figure*}
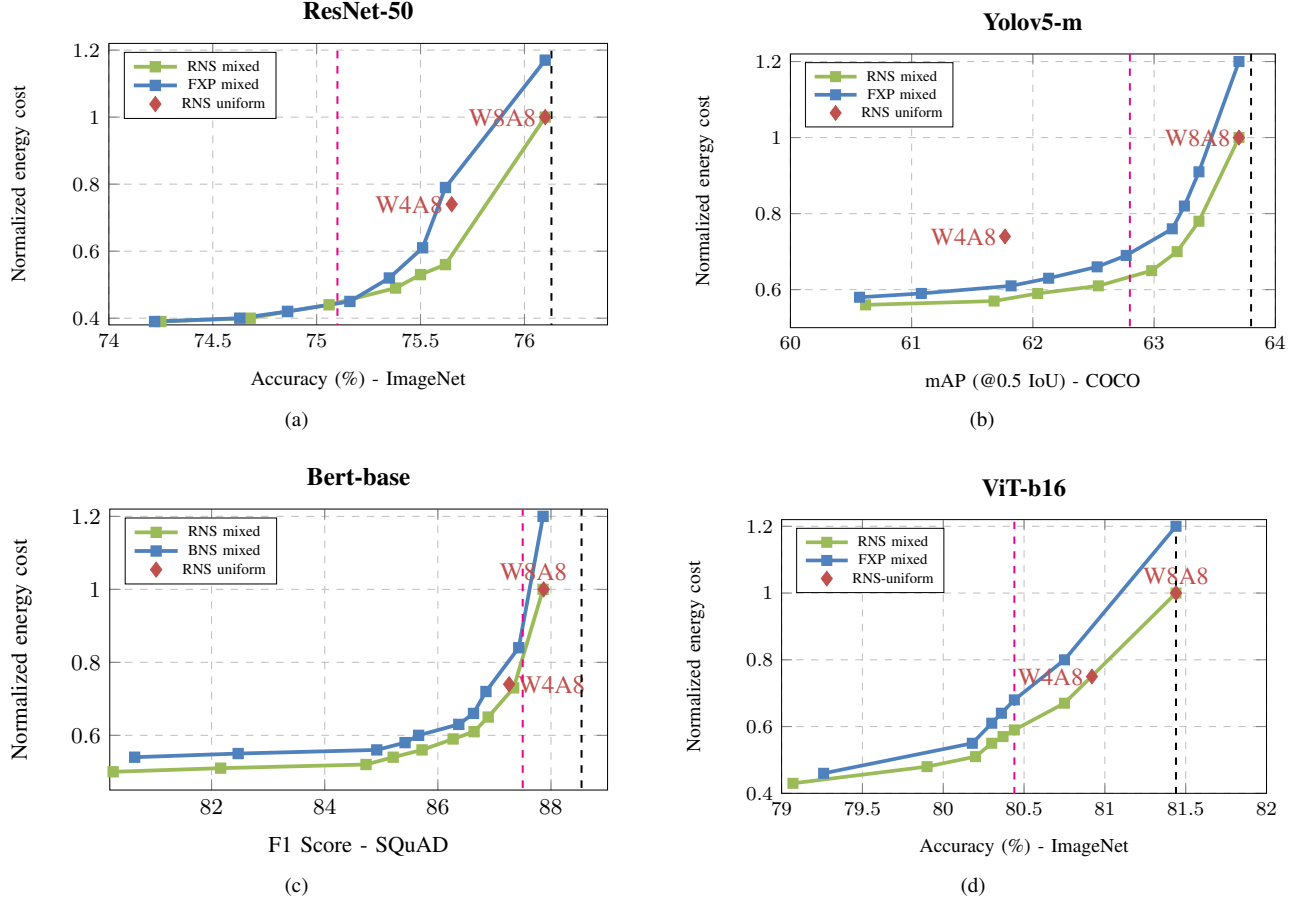

The goal of the dynamic quantization optimization methodology is to find the best-performing quantization configuration for a given accuracy degradation constraint. To explore the trade-off between accuracy and energy efficiency, we obtain optimal quantization configurations for various loss-function degradation constraints and plot the normalized cost versus the model's resulting performance (accuracy, mAP, or F1 score). The cost is estimated using Equation~\ref{eq:metric}, with power consumption values $P_j$, $P_\text{mem}$, and $P_\text{ppu}$ obtained from post-PnR power analysis. We evaluate the proposed methodology and hardware architecture using four popular NN models: ResNet-50 \cite{resnet} and VisionT Transformer (ViT-16)~\cite{vit} for image classification (on ImageNet), Yolov5-m \cite{yolo} for image segmentation (on the COCO dataset), and Bert-base \cite{bert} for language understanding (on the SQuAD dataset).  We compare the RNS mixed-precision implementation against a FXP system that supports power-of-2-bit operands (W8A8, W4A8, WA4A), as this is common among FXP ASIC accelerators \cite{bitblade,diana,bitfusion}. The power efficiency of the RNS-based system during W8A8 uniform quantization is used as the baseline. Results are visualized in Fig.~\ref{fig:performance}, where the performance-cost trade-off curves for RNS and FXP are plotted.

 The end-to-end improvements achieved by the RNS system are due to the more efficient implementation of the TPAs, which account for about $40-50\%$ of the total power consumption. We note, however, that plots of Figure~\ref{fig:performance} refer to total power consumption, according to Equation~\ref{eq:metric}, and thus \emph{include memory power consumption} as well, which is the same for both FXP and RNS systems.

\begin{itemize}
    \item \textbf{ResNet}: The RNS-based accelerator achieves a $1.20\times$ end-to-end power efficiency increase compared to the FXP counterpart, for lossless (zero accuracy drop) inference. It maintains superior performance within the 1\% accuracy drop mark. RNS and FXP are practically equivalent for low accuracy constraints. 

    \item \textbf{Yolov5-m}: The RNS-based accelerator achieves a $1.20\times$ end-to-end power efficiency increase compared to the FXP counterpart, for lossless inference. It maintains a distinctly superior performance for \emph{all accuracy drop constraints} of $\approx 1.10\times$. 

    \item  \textbf{Bert-base}: The RNS-based accelerator achieves a $1.20\times$ end-to-end power efficiency increase compared to the FXP counterpart, for lossless inference. It maintains a superior performance for all accuracy drop constraints.

    \item  \textbf{ViT}: RNS achieves $1.15$--$1.20 \times$ performance improvement within the 1\% accuracy drop mark, while maintaining a superior performance for all accuracy constraints.
\end{itemize}

\begin{figure}[]
    \centering
    \includegraphics[width=.4\textwidth]{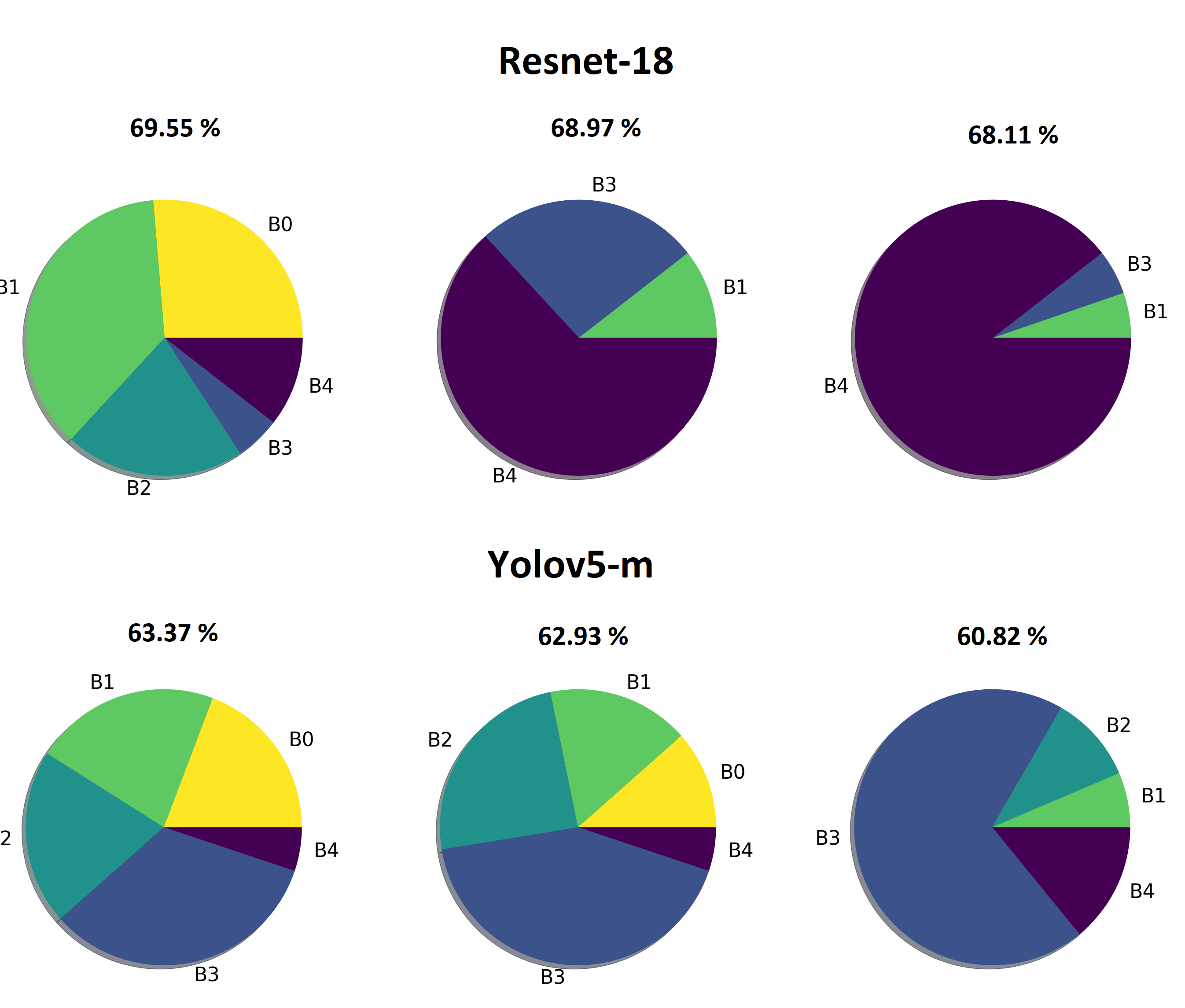}
    \caption[Base utilization frequency]{Base utilization frequency for various accuracy levels, for ResNet-18 (top) and Yolov5-m (bottom). The Yolov5 model is more sensitive to quantization, thus a lower frequency of small bases ($\mathcal{B}_4$) is observed.}
    \label{fig:base_usage}
\end{figure}

RNS benefits are greater in models that exhibit higher sensitivity to various quantization configurations, such as the Yolov5 model. The increased sensitivity forces the mixture of the utilized configurations towards quantizations with more bits, or equivalently, larger bases. This favors RNS, which performs large-bit-width operations more efficiently. The resulting base utilization frequency for various accuracy levels and models is shown in Figure~\ref{fig:base_usage}. Using a $0.3\%$ accuracy drop constraint for ResNet-18, $\mathcal{B}_0$ utilization is $\approx 25\%$ among all layers, while for a $2\%$ constraint $\mathcal{B}_0$ is not utilized at all, and $\mathcal{B}_4$ is by far the most frequent. In contrast, $\mathcal{B}_4$ frequency is considerably smaller in Yolov5-m, due to the increased sensitivity of this model to quantization error.

\subsection{Comparison to RISC-V-based accelerators}

Tightly coupling RISC-V processors with DNN accelerators is a common approach that benefits from the mature RISC-V software ecosystem and programming flexibility.
MX~\cite{mx} introduces a matrix extension to the RISC-V ISA built on top of the RISC-V Vector Extension (RVV). The architecture achieves high PE utilization for dense matrix multiplication, reaching $97.2\%$ with two cores on a $64{\times}64{\times}64$ workload. However, utilization decreases with higher degrees of parallelism, dropping to $78.7\%$ in a 64-core configuration.
Spatz~\cite{spatz} improves the energy efficiency of RISC-V vector processors through optimized vector register-file sizing. It achieves utilization levels of up to $97.9\%$ for matrix multiplication and $95\%$ for convolution workloads, although at substantially lower parallelism levels (approximately 15 operations/cycle) than the proposed architecture.
Quadrilatero~\cite{quadrilatero} proposes a programmable systolic-array coprocessor implementing a RISC-V matrix extension. The design achieves up to $99.4\%$ utilization for large matrices, depending on the data format, but is restricted to matrix-multiplication applications.

Dynamic-precision RISC-V-based accelerators have also been recently developed.
MARSELLUS~\cite{marsellus}, a precision-scalable FXP accelerator integrated into a RISC-V SoC, supports precisions of $2$--$8$ bits. It utilizes a bit-serial dataflow and achieves a maximum energy efficiency of $12.4$ TOPS/W for W2A2 quantization and $7.8$ TOPS/W for W4A4. The proposed RNS-based accelerator achieves $9.96$ TOPS/W for W4A4 quantization, marking a $1.27\times$ improvement. Flex-v~\cite{flexv} is another RISC-V accelerator based on the XPulpNN platform~\cite{xpulpnn} that supports dynamic bit-scalable execution using a custom mixed-precision dot-product unit and introduces DNN-tailored instructions to increase MAC unit utilization.
Implemented on the same 22~nm technology, it achieves an energy efficiency of $3.26$ and $0.87$ TOPS/W for W2A2 and W8A8 quantization, respectively. The proposed accelerator is $5.88\times$ more power efficient.

A RISC-V-integrated accelerator targeting transformer models on edge devices is proposed in \cite{riscvattent}. The system supports 8-bit quantized attention and achieves $2.96$ TOPS/W for MobileBert (a scaled-down version of Bert), implemented on a 22 nm technology. Our custom-ISA RNS-based accelerator, implemented on the same node, achieves an average of $4.8$~TOPS/W when executing a Bert transformer block (assuming a scaled memory system that can fit the weight and activation tensors).

Overall, the proposed architecture achieves superior end-to-end power efficiency, with utilization levels and workload coverage similar to or higher than those of the RISC-V-based counterparts. This is attributed to the efficiency of the ML-tailored ISA and the lightweight control cores, combined with the dynamic-precision execution scheme and RNS. End-to-end quantitative comparisons are reported in Table~\ref{t:mxp_sota}.

\begin{table}[t]
\caption{Comparison with state-of-the-art dynamic-precision systems}
\centering
\resizebox{!}{0.20\linewidth}{
\footnotesize
\setlength{\tabcolsep}{2pt}
\begin{tabular}{@{}lccccc@{}}
\toprule
 &
~\cite{bitblade} &
~\cite{diana}$^\dagger$ &
~\cite{marsellus} &
~\cite{flexv} &
\textbf{This work} \\
\midrule
Process      & 28nm & 22nm & 22nm & 22nm & \textbf{22nm} \\
Freq. (MHz)  & 44--195 & 50--320 & 420 & 463 & \textbf{800} \\
Area (mm$^2$)& 0.71 & 6.61 & 1.9 & 0.53 & \textbf{2.25} \\
SRAM (kB)    & 144 & 896 & 128 & 128 & \textbf{344} \\
Precision    & 2/4/8 & 2/4/8 & 2--8 & 2/4/8 &
\textbf{W:3--8} \newline \textbf{A:4/8} \\
\midrule
\begin{tabular}[c]{@{}c@{}} Power Eff.\\ (TOPS/W)$^*$ \end{tabular}
&
\begin{tabular}[c]{@{}c@{}}
3 @W8A8$^{**}$\\
12 @W4A4$^{**}$\\
44.1 @W2A2
\end{tabular}
&
4.1
&
\begin{tabular}[c]{@{}c@{}}
7.8 @W4A4\\
12.4 @W2A2
\end{tabular}
&
\begin{tabular}[c]{@{}c@{}}
0.87 @W8A8\\
3.26 @W2A2
\end{tabular}
&
\begin{tabular}[c]{@{}c@{}}
\textbf{5.14 @W8A8}\\
\textbf{8.22 @W4A7}\\
\textbf{9.95 @W4A4}\\
\textbf{10.47 @W3A4}
\end{tabular}
\\
\bottomrule
\multicolumn{2}{l}{$^*$Peak eff., 1 MAC = 2 OPS}& \multicolumn{2}{l}{$^{**}$ data points from plots } & \multicolumn{2}{l}{$^\dagger$ digital subsystem}
\end{tabular}}
\label{t:mxp_sota}
\end{table}

\subsection{Comparison to fixed-function accelerators}

We also compare the RNS MXP accelerator with state-of-the-art dynamic-precision fixed-function accelerators. Results are reported in Table~\ref{t:mxp_sota}. Supported bit precisions (`W' for weights and `A' for activations) are indicated for each system.  DIANA \cite{diana}, a hybrid digital and analog DNN accelerator, supports 2-, 4- and 8-bit operands and achieves a maximum energy efficiency of $4.1$ TOPS/W using the digital sub-system. A $2.54\times$ improvement in maximum energy efficiency is achieved over this system. BitBlade~\cite {bitblade} utilizes bit-wise serial computations to efficiently support 2-,~4-~and 8-bit precisions, reaching a maximum energy efficiency of $44.1$~TOPS/W for W2A2 quantization. However, as this system is optimized for low precision, energy efficiency significantly drops to $\approx 3$ TOPS/W for W8A8 quantization. For this configuration, the proposed accelerator outperforms BitBlade by $1.70\times$.

DataMaestro~\cite{datamaestro} is a data streaming engine that aims to bridge the gap between the flexibility of ISA-based systems and the performance of fixed-function accelerators. While not fully programmable like the proposed ISA-based architecture (it requires a host RISC-V system), it supports similar features such as decoupled memory access and execution, and fine-grained addressing modes, albeit without dynamic-precision capabilities. It achieves very high utilization ($\approx 95$ -- $99\%$) across diverse NN workloads; however, the data streaming engines consume a considerable $15\%$ of the total power (versus the $4.1\%$ overhead that the proposed mCores introduce), leading to a maximum energy efficiency of $2.57$ TOPS/W when implemented on a $22$~nm FDX technology node.

\section{Conclusion}

In this work, we presented a machine-learning-oriented ISA and reconfigurable accelerator architecture that bridges the gap between highly efficient fixed-function DNN accelerators and flexible ISA-based solutions. The proposed architecture employs lightweight control cores, decoupled execution domains, and dynamic-precision SIMD processing to sustain high utilization across diverse DNN workloads while incurring only a small control overhead. Its data-representation-independent ISA enables systematic evaluation and deployment of alternative numerical formats.  Using an RNS-based implementation as a case study, we demonstrated that DP RNS arithmetic can improve the accuracy--efficiency trade-off of modern DNNs, yielding up to $1.2\times$ higher end-to-end energy efficiency than a mixed-precision fixed-point counterpart. A complete 22-nm implementation achieves up to $10.47$~TOPS/W and compares favorably with state-of-the-art RISC-V-based and fixed-function accelerators. These results demonstrate that ML-tailored programmable architectures, combined with dynamic precision and alternative numerical representations, offer a promising path toward efficient, adaptable next-generation AI accelerators.

\IEEEtriggeratref{4}
\IEEEtriggercmd{\balance}
\bibliographystyle{IEEEtran}
\bibliography{references}

\end{document}

%% file: preamble.tex
\usepackage{ifthen}
\usetikzlibrary{shadows.blur}
\usetikzlibrary{positioning}
\usetikzlibrary{calc,intersections,through,backgrounds}
\def\myarm{1cm}
\def\myangle{0}
\tikzset{
  arm/.default=1cm,
  arm/.code={\def\myarm{#1}}, 
  angle/.default=0,
  angle/.code={\def\myangle{#1}} 
}
\tikzset{
    myncbar/.style = {to path={
        let
            \p1=($(\tikztotarget)+(\myangle:\myarm)$)
        in
            -- ++(\myangle:\myarm) coordinate (tmp)
            -- ($(\tikztotarget)!(tmp)!(\p1)$)
            -- (\tikztotarget)\tikztonodes
    }}
}
\tikzset{%
 blockPE/.style    = {draw, fill = yellow!30!white,thick,node distance = 0.1cm, minimum height = 0.1em,
    minimum width = 0.1em},
  blocknew/.style    = {draw, fill = red!20!white,thick, rectangle,node distance = 1.5cm, minimum height = 1em,
    minimum width = 5em},
    blocknewdata/.style    = {draw, fill = green!20!white,thick, rectangle,node distance = 1.5cm, minimum height = 1em,
    minimum width = 5em},
    Lut/.style    = {draw, thick,fill = blue!20!white, rectangle,node distance = 0.5cm, minimum height = 1.2em,
    minimum width = 0.5em},
     blocknewtonos/.style    = {draw, thick,fill = green!20!white, rectangle,node distance = 1.5cm, minimum height = 6em,
    minimum width = 3em},
      blocknewthin/.style    = {draw, thick,fill = black!10!white, rectangle,node distance = 1.5cm, minimum height = 0.5em,
    minimum width = 9em},
    blocknewthin1/.style    = {draw, thick,fill = red!20!white, rectangle,node distance = 1.0cm, minimum height = 0.5em,
    minimum width = 5em},
     blocknewtonos2/.style    = {draw, thick,fill = green!20!white, rectangle,node distance = 1.0cm, minimum height = 6em, inner sep=0pt,minimum width = 1.5em},
    bullet/.style      = {draw, circle, minimum size=0.2em, inner sep=0pt,node distance = 1.5cm}, 
         bulletw/.style      = {draw, circle,fill = black, minimum size=0.4em, inner sep=0pt,node distance = 1.5cm}, 
  sum/.style      = {draw, rectangle, node distance = 1.5cm}, 
   mul/.style      = {draw, circle, minimum size=0.8em,inner sep=1pt,node distance = 1.5cm}, 
   sum1/.style ={rectangle,draw,minimum width=2cm} ,
  input/.style    = {coordinate}, 
   output/.style   = {coordinate}, 
   multiplexer/.style={
    draw,
    trapezium,
    node distance = 0.8cm,
    minimum height = 3em,
    minimum width = 13em,
    inner ysep=6pt,
    inner xsep=6pt,
    shape border uses incircle,
    shape border rotate=270,
    minimum size=8pt
  }
}

\usetikzlibrary{shapes,arrows} 

\def\myarm{1cm}
\def\myangle{0}
\tikzset{
  arm/.default=1cm,
  arm/.code={\def\myarm{#1}}, 
  angle/.default=0,
  angle/.code={\def\myangle{#1}} 
}

\tikzset{
    myncbar/.style = {to path={
        let
            \p1=($(\tikztotarget)+(\myangle:\myarm)$)
        in
            -- ++(\myangle:\myarm) coordinate (tmp)
            -- ($(\tikztotarget)!(tmp)!(\p1)$)
            -- (\tikztotarget)\tikztonodes
    }}
}
\tikzstyle{block} = [draw,circle,fill=blue!10,inner sep=-2.5,outer sep=0,font=\fontsize{8}{0}\selectfont,scale=1.6]
\tikzstyle{shiftblock} = [draw,circle,fill=red!05,inner sep=1,outer sep=0]

\tikzstyle{adder} =  [draw, rectangle, node distance = 2cm,fill=red!15,inner sep=2,outer sep=0,scale=1.2]

\tikzstyle{branch}=[fill,shape=circle,minimum size=4pt,inner sep=0pt]

\makeatletter
\def\pgfpie@legend#1{%
  \coordinate[xshift=0.4cm, 
  yshift={(\the\pgfpie@sliceLength*0.5-1)*0.5cm}] (pgfpie@legendpos) at
  (current bounding box.east); 

  \scope[node distance=0.3cm] 
    \foreach \pgfpie@p/\pgfpie@t [count=\pgfpie@i from 0] in {#1}
    {
      \pgfpie@findColor{\pgfpie@i}
      \node[draw, fill={\pgfpie@thecolor}, \pgfpie@style, below of={pgfpie@legendpos}, label={0:{\pgfpie@t}}] (pgfpie@legendpos) {};
    }
  \endscope
}
\makeatother

%% file: Figures/conv_mem_layout.tex
\begin{tikzpicture}
[auto, thick, node distance=2cm, >=triangle 45]


   \foreach \x [evaluate=\x as \y using \x-4] in {5,5.5} {    

\fill[black!6,fill opacity=1.0, ] (\y,\y) rectangle (\x,\x);
        \draw[step=5mm, thin, black!2] (\y,\y) grid (\x,\x); 
      \draw[black, dotted] (\y,\y) rectangle (\x,\x);
}

  \fill[black!20,fill opacity=1.0] (2,2) rectangle (6,6);
        \draw[step=5mm, thin, black!30] (2,2) grid (6,6); 
        \draw[black,dotted ] (2,2) rectangle (6,6);
\fill[blue!50,opacity=0.8] (3,5) rectangle (5,5.5);
\fill[blue!50,opacity=0.8] (2.5,2.5) rectangle (3.0,5.5);

\fill[red!60,opacity=0.8] (3,3) rectangle (5,5);
\fill[magenta!50,opacity=0.8] (3,2.5) rectangle (5.5,3);
\fill[magenta!50,opacity=0.8] (5.0,3) rectangle (5.5,5.5);


\foreach \x [evaluate=\x as \y using \x+2] in {8,11} {    
\foreach \k [evaluate=\k as \l using \k+1.5] in {-1.0,2,4.5} {

\fill[white,fill opacity=0.8, general shadow={fill=black!40,shadow scale=1.0}] (\x,\k) rectangle (\y,\l);
        \draw[step=5mm, thin, black!20] (\x,\k) grid (\y,\l); 
        \draw[black!50, dotted] (\x,\k) rectangle (\y,\l);
}
}

\fill[red!60,opacity=0.8] (8,5) rectangle (10,5.5);
\fill[blue!50,opacity=0.8] (9.5,5.5) rectangle (10,6);
\fill[magenta!50,opacity=0.8] (8,4.5) rectangle (8.5,5);

\fill[red!60,opacity=0.8] (11,5) rectangle (13,5.5);
\fill[blue!50,opacity=0.8] (12.5,5.5) rectangle (13,6);
\fill[magenta!50,opacity=0.8] (11,4.5) rectangle (11.5,5);

\fill[magenta!50,opacity=0.8] (8,2.5) rectangle (10,3);
\fill[blue!50,opacity=0.8] (9.5,3) rectangle (10,3.5);
\fill[magenta!50,opacity=0.8] (8,2) rectangle (8.5,2.5);

\fill[blue!50,opacity=0.8] (8,0) rectangle (10,0.5);

\fill[blue!50,opacity=0.8] (11,-1.0) rectangle (11.5,0.5);

\draw[color=black] (9.0,5.8) node{$(4n)_1$};
\draw[color=black] (9.0,5.3) node{$(4n)_2$};
\draw[color=black] (9.0,4.8) node{$(4n)_3$};
 
 \draw[color=black] (12.0,5.8) node{$(4n+1)_1$};  
 \draw[color=black] (12.0,5.3) node{$(4n+1)_2$};  
 \draw[color=black] (12.0,4.8) node{$(4n+1)_3$};  

 \draw[color=black] (9,3.3) node{$(4n+4)_1$};  
 \draw[color=black] (9,2.8) node{$(4n+4)_2$};  
   \draw[color=black] (9,2.3) node{$(4n+4)_3$};

\draw[green] (1.0,4.5) rectangle (3,5);
\draw[green] (3.0,4.5) rectangle (5,5);
\draw[green] (5.0,4.5) rectangle (7,5);

 \draw[color=black] (2.1,4.8) node[] {$(4n)_1$};  
 \draw[color=black] (4.1,4.8) node[] {$(4n)_2$};
 \draw[color=black] (6.1,4.8) node[] {$(4n)_3$};

\draw[green] (1.0,4) rectangle (3,4.5);
\draw[green] (3.0,4) rectangle (5,4.5);
\draw[green] (5.0,4) rectangle (7,4.5);

\draw[color=black] (2.1,4.3) node[] {$(4n+1)_1$};  
 \draw[color=black] (4.1,4.3) node[] {$(4n+1)_2$};
 \draw[color=black] (6.1,4.3) node[] {$(4n+1)_3$};

\draw[green] (1.0,3.5) rectangle (3,4);
\draw[green] (3.0,3.5) rectangle (5,4);
\draw[green] (5.0,3.5) rectangle (7,4);

   \draw[color=black] (2.1,3.8) node[] {$(4n+2)_1$};  
 \draw[color=black] (4.1,3.8) node[] {$(4n+2)_2$};
 \draw[color=black] (6.1,3.8) node[] {$(4n+2)_3$};

\draw[green] (1.0,3) rectangle (3,3.5);
\draw[green] (3.0,3) rectangle (5,3.5);
\draw[green] (5.0,3) rectangle (7,3.5);

\draw[color=black] (2.1,3.3) node[] {$(4n+3)_1$};  
 \draw[color=black] (4.1,3.3) node[] {$(4n+3)_2$};
 \draw[color=black] (6.1,3.3) node[] {$(4n+3)_3$};
 
\draw[green] (1.0,2.5) rectangle (3,3);
\draw[green] (3.0,2.5) rectangle (5,3);
\draw[green] (5.0,2.5) rectangle (7,3);

\draw[color=black] (2.1,2.8) node[] {$(4n+4)_1$};  
 \draw[color=black] (4.1,2.8) node[] {$(4n+4)_2$};
 \draw[color=black] (6.1,2.8) node[] {$(4n+4)_3$};

 \draw[color=blue] (3.9,6.4) node{Feature Map Block};
 \draw[color=blue] (10.8,6.8) node{Feature Map Memory (OP1MEM)};
 \draw[color=black] (9.2,6.4) node{Bank 0};
 \draw[color=black] (12.2,6.4) node{Bank 1};

 \draw[color=black, scale=1.2] (7.5,3.5) node{$\vdots$};
\draw[color=black, scale=1.2] (10.0,3.5) node{$\vdots$};
\draw[color=black, scale=1.2] (11.3,3.3) node{$\ldots$};
 
\draw[color=blue] (10.8,1.2) node{Border Cache};
\draw[color=black] (9.2,0.8) node{Row cache (bank 0)};
\draw[color=black] (12.4,0.8) node{Col. cache (bank 1)};

\draw[black!60] (7.6,1.7) rectangle (14,7.1);
\draw[black!60] (7.6,-1.2) rectangle (14,1.5);

\draw[->] (3.3,5.4) .. controls (3.0,1.7) .. (9,0.4);

\draw[->  ] (2.7,4.2) .. controls (3.5,-0.3) .. (11.1,-0.1);

\fill[blue!60,opacity=0.8] (-1.0,7.2) rectangle (-0.6,7.6);
\fill[red!50,opacity=0.8] (3.6,7.2) rectangle (4,7.6);
\fill[magenta!50,opacity=0.8]  (9.1,7.2) rectangle (9.5,7.6);
\draw[color=blue!60] (1.5,7.4) node{Load from Border Buffer};
\draw[color=red!60] (6.5,7.4) node{Load from current FMEM addr.};
\draw[color=magenta!60] (12,7.4) node{Pre-Fetch from next FMEM addr.};

\end{tikzpicture} 

%% file: Figures/util_matmul_conv.tex
\definecolor{bblue}{HTML}{4F81BD}
\definecolor{rred}{HTML}{C0504D}
\definecolor{ggreen}{HTML}{9BBB59}

\setlength{\tabcolsep}{-1em}
\begin{tabular}{@{\hskip2pt}l@{\hskip0pt}l}
 \begin{tikzpicture}[baseline=0]
        \begin{axis}[
        xlabel={$K$},
        height=6cm,
        width=5cm,
            every tick label/.append style={font=\footnotesize},
    every axis label/.append style={font=\footnotesize},
        grid=major,
        legend pos=south east,
        domain=32:512,
        legend style={nodes={scale=0.6, transform shape}}
    ]
    \addplot[bblue, ultra thick] {(64*64*x/512)/(64/16*(3+64/32*(5+x)))}; 
    \addlegendentry{$(N,M){=}(64,64)$)}
     \addplot[rred, ultra thick] {(128*128*x/512)/(128/16*(3+128/32*(5+x)))}; 
       \addlegendentry{$(N,M){=}(128,128)$)}
      \addplot[ggreen, ultra thick] {(512*512*x/512)/(512/16*(3+512/32*(5+x)))}; 
        \addlegendentry{$(N,M){=}(512,512)$)}
    \end{axis}
  \end{tikzpicture}
  &
  \begin{tikzpicture}[baseline=0]
     \begin{axis}[
        xlabel={input channels},
          height=6cm,
        width=5cm,
        grid=major,
        legend pos=south east,
        domain=32:512,
            every tick label/.append style={font=\footnotesize},
    every axis label/.append style={font=\footnotesize},
        legend style={nodes={scale=0.6, transform shape}}
    ]
     \addplot[bblue, ultra thick] {(64/4*64/4*64/32*x*9)/(64/4*(6+64/4*(8+10+64/32*(8+x*9))))};
    \addlegendentry{3x3 kernel}
     \addplot[rred, ultra thick] {(64/4*64/4*64/32*x*25)/(64/4*(6+64/4*(8+10+64/32*(8+x*25))))};
       \addlegendentry{5x5 kernel}
      \addplot[ggreen, ultra thick] {(64/4*64/4*64/32*x*49)/(64/4*(6+64/4*(8+10+64/32*(8+x*49))))};
        \addlegendentry{7x7 kernel}
    \end{axis}
  \end{tikzpicture}

  \end{tabular}

%% file: Figures/bert_f1.tex
\begin{tikzpicture}
\centering
\begin{axis}[
   ybar,
    width = 8cm,
    height = 5cm,
    ylabel=F1 Score,
    ymin=87,
    ymax=88.6,
    xmax=4,
    ymajorgrids=true,
    xmajorticks=false,
   enlargelimits=0.3,
nodes near coords,
every node near coord/.append style={font=\footnotesize},
legend style={nodes={scale=0.6, transform shape}},
 legend pos=north east,
    grid style=dashed,
    legend image code/.code={
        \draw [#1] (0cm,-0.1cm) rectangle (0.3cm,0.1cm); }]
     \addplot[fill=blue!100] coordinates {(0,88.54)};
     \addlegendentry{FLP32}
    \addplot[fill=blue!70] coordinates {(1,87.77)};
    \addlegendentry{FXP W8A8}
    \addplot[fill=blue!40] coordinates {(2,87.74)};
    \addlegendentry{FXP W8A8 + approx. softmax }
    \addplot[fill=blue!10] coordinates {(3,87.65)};
    \addlegendentry{FXP W8A8 + approx. (softmax+GELU) }

\end{axis}
\end{tikzpicture}

%% file: Figures/util_dw_pw.tex
\definecolor{bblue}{HTML}{4F81BD}
\definecolor{ggreen}{HTML}{9BBB59}

\begin{tikzpicture}
    \begin{axis}[
        xlabel={output channels},
        ylabel={Utilization},
         height=6cm,
         ylabel style={yshift=-1em},
        grid=major,
        legend pos=south east,
        domain=32:512,
        legend style={nodes={scale=0.8, transform shape}}
    ]
         \addplot[bblue, ultra thick] { ((9+x)/512)/(max(9/16,x/496))};
        \addlegendentry{$(N_{\text{TPA}},N_\text{PE})=(16,32)$}
         \addplot[ggreen, ultra thick] { ((9+x)/256)/(max(9/16,x/240))};
        \addlegendentry{$(N_{\text{TPA}},N_\text{PE})=(16,16)$}
    \end{axis}
\end{tikzpicture}

%% file: Figures/base_comp_rn18.tex
  \begin{tikzpicture}[spy using outlines={
            rectangle, 
            blue, 
            magnification=5,
            height=3cm,
            width=1.5cm,
            connect spies}]
    \begin{axis}[%
      ylabel=Normalized Energy Cost,
      xlabel=$\delta L$,
       width = 0.5\textwidth,
     height = 9cm,
    legend pos=north east,
    enlarge y limits=0.05,
    every tick label/.append style={font=\small},
    every axis label/.append style={font=\small},
     every axis plot/.append style={thick},
    ymajorgrids=true,
    xmajorgrids=true,
    grid style=dashed,
    legend style={nodes={scale=0.7, transform shape}}
    ]
      
     \addplot[
    color=orange,
    mark=square*,
    mark size=1pt
    ]
    coordinates {
 (30,0.8430347439849781)(50,0.8041069283100709)(100,0.743050251927944)(200,0.6430657598709099)(300,0.5710795260933552)(400,0.5197887783393316)(600,0.4599882167628781)(800,0.42818514908523125)};
\addlegendentry{$\{5, 7, 9, 16, 31\}$} 

  \addplot[
    color=red,
    mark=square*,
    mark size=1pt
    ]
    coordinates {
(30,0.8387720906791031)(50,0.7782014316670597)(100,0.6974459774050743)(200,0.5884890361580012)(300,0.5213583133591773)(400,0.48491728548838203)(600,0.454783359678355)(800,0.4473861421557018)
};
\addlegendentry{$\{5, 7, 9, 31, 32\}$} 

  \addplot[
    color=green,
    mark=square*,
    mark size=1pt
    ]
    coordinates {
(30,0.8674567587796728)(50,0.8169924513629345)(100,0.7394484218059023)(200,0.6304574978033404)(300,0.5570380221029508)(400,0.5045319734481236)(600,0.4498475042448039)(800,0.4219453527542285)
};
\addlegendentry{$\{3, 5, 7, 11, 13, 16\}$} 

 \addplot[
    color=magenta,
    mark=square*,
    mark size=1pt
    ]
    coordinates {
(30,0.878783971139412)(50,0.8269546327172431)(100,0.7541544944454366)(200,0.6345456228863414)(300,0.5538765878939969)(400,0.5015787927419374)(600,0.4470007816510535)(800,0.4151341738242187)
};
\addlegendentry{$\{3, 5, 7, 11, 13, 32\}$} 

 \addplot[
    color=blue,
    mark=square*,
    mark size=1pt
    ]
    coordinates {
(30,0.8806612297258897)(50,0.8241740411208558)(100,0.7461649301780616)(200,0.631565053213733)(300,0.5551325860310412)(400,0.5026009547620132)(600,0.4499484024936094)(800,0.4204591392507709)
};
\addlegendentry{$\{3, 5, 7, 11, 16, 31\}$}

 \addplot[
    color=cyan,
    mark=square*,
    mark size=1pt
    ]
    coordinates {
(30,0.8495459799389935)(50,0.7818920265004907)(100,0.7140311230294201)(200,0.6174685097918331)(300,0.5523261704370828)(400,0.5061879438298347)(600,0.45593975901373757)(800,0.42277217528185024)
};
\addlegendentry{$\{5, 7, 9, 11, 13, 16\}$} 
\coordinate (sp) at (axis cs:800,0.43);
\spy [
spy connection path={\draw[->] (tikzspyonnode) -- (tikzspyinnode);}]
on (sp) in node [above, shift={(0.0cm,0.5cm)}];
\end{axis}
 \end{tikzpicture}

%% file: Figures/base_comp_rn50.tex
  \begin{tikzpicture}
    \begin{axis}[%
      ylabel=Normalized Energy Cost,
      xlabel=$\delta L$,
       width = 0.5\textwidth,
     height = 9cm,
    legend pos=north east,
    enlarge y limits=0.05,
    every tick label/.append style={font=\small},
    every axis label/.append style={font=\small},
     every axis plot/.append style={thick},
    ymajorgrids=true,
    xmajorgrids=true,
    grid style=dashed,
    legend style={nodes={scale=0.7, transform shape}}
    ]
      
  \addplot[
    color=red,
    mark=square*,
    mark size=2pt
    ]
    coordinates {
(10,0.6654741147110504)(20,0.6216374380751617)(30,0.592803059282191)(50,0.5685864020147636)(100,0.5280775854146299)(200,0.4820487168793239)(400,0.44598510143841874)(500,0.44173250577297535)
};
\addlegendentry{$\{5, 7, 9, 31, 32\}$} 

 \addplot[
    color=magenta,
    mark=square*,
    mark size=2pt
    ]
    coordinates {
(10,0.6850286834082887)(20,0.6353475612100512)(30,0.6064036912327609)(50,0.5715796084583386)(100,0.5194739516959076)(200,0.4635506977239717)(400,0.4110737384993934)(500,0.3996913642588289)
};
\addlegendentry{$\{3, 5, 7, 11, 13, 32\}$} 

 \addplot[
    color=cyan,
    mark=square*,
    mark size=2pt
    ]
    coordinates {
(10,0.6800110569754462)(20,0.6295223105470537)(30,0.600493776094731)(50,0.5669254494490417)(100,0.5217292655810237)(200,0.4717262795158251)(400,0.42049369937335934)(500,0.40738547184372276)
};
\addlegendentry{$\{5, 7, 9, 11, 13, 16\}$}

\end{axis}
 \end{tikzpicture}

%% file: Figures/base_comp_yolo.tex
  \begin{tikzpicture}
    \begin{axis}[%
      ylabel=Normalized Energy Cost,
      xlabel=$\delta L$,
       width = 0.5\textwidth,
     height = 9cm,
    legend pos=north east,
    enlarge y limits=0.05,
    every tick label/.append style={font=\small},
    every axis label/.append style={font=\small},
     every axis plot/.append style={thick},
    ymajorgrids=true,
    xmajorgrids=true,
    grid style=dashed,
    legend style={nodes={scale=0.7, transform shape}}
    ]
      
  \addplot[
    color=red,
    mark=square*,
    mark size=2pt
    ]
    coordinates {
(50,4.43564574619349)(100,4.255104091118413)(200,3.974441328888573)(400,3.736817217373436)(700,3.6114381472769623)(1000,3.5441987791902103)(1500,3.4563886246664626)(2500,3.3490928081953637)(4000,3.270528282788461)

};
\addlegendentry{$\{5, 7, 9, 31, 32\}$}

\end{axis}
 \end{tikzpicture}

%% file: Figures/power.tex

\definecolor{bblue}{HTML}{4F81BD}
\definecolor{rred}{HTML}{C0504D}
\definecolor{ggreen}{HTML}{9BBB59}
\definecolor{mmagenta}{HTML}{C27BA0} 
\definecolor{cturq}{HTML}{2CA6A4}
\usetikzlibrary{arrows.meta}

\begin{tikzpicture}
\begin{axis}[
    ybar stacked,
    bar width=16pt,
    width=0.9\columnwidth,
    height=0.65\columnwidth,
    ymin=0,
    ymax=105,
    ylabel={Power (mW)},
    ylabel style={yshift=-1em},
    symbolic x coords={W8A8,W4A7,W3A4},
    xtick=data,
    enlarge x limits=0.25,
    ymajorgrids=true,
    grid style={dashed,gray!30},
    axis line style={black},
    tick label style={font=\footnotesize},
    label style={font=\footnotesize},
    legend style={
        at={(0.5,1.03)},
        anchor=south,
        legend columns=3,
        draw=none,
        font=\footnotesize
    }
]

\addplot+[fill=bblue, draw=black] coordinates {
    (W8A8,50.40)
    (W4A7,24.16)
    (W3A4,19.36)
};

\addplot+[fill=rred, draw=black] coordinates {
    (W8A8,7.22)
    (W4A7,5.12)
    (W3A4,3.65)
};

\addplot+[fill=mmagenta, draw=black] coordinates {
    (W8A8,4.02)
    (W4A7,4.02)
    (W3A4,4.24)
};

\addplot+[fill=ggreen, draw=black] coordinates {
    (W8A8,22.21)
    (W4A7,16.62)
    (W3A4,11.55)
};

\addplot+[fill=cturq, draw=black] coordinates {
    (W8A8,15.10)
    (W4A7,12.50)
    (W3A4,9.90)
};

\legend{Tensor SIMD, PPUs + conversion, Control, Memory, Clock}

\draw[-{Latex[length=2mm]}, thick]
    (axis cs:W8A8,99) to[bend left=20]
    node[midway, above, font=\footnotesize] {$-36.4\%$}
    (axis cs:W4A7,63);

\draw[-{Latex[length=2mm]}, thick]
    (axis cs:W4A7,65) to[bend left=20]
    node[midway, above, font=\footnotesize] {$-25.4\%$}
    (axis cs:W3A4,50);

\end{axis}
\end{tikzpicture}

%% file: Figures/area.tex
\definecolor{bblue}{HTML}{4F81BD}
\definecolor{rred}{HTML}{C0504D}
\definecolor{ggreen}{HTML}{9BBB59}
\definecolor{mmagenta}{HTML}{C27BA0}

\begin{tikzpicture}[baseline=0]
 \pie[font=\small,
   draw=white,
   color={ggreen, ggreen!70, ggreen!40, rred,  bblue, mmagenta},
   every only number node/.style={text=black}]{25.4/OP1MEM, 41.9/OP2MEM,  5.25/OP1CACHE \& OP1FUSED,  7/PPUs \& conversion,  15.4/TPAs, 5.0/control}
\end{tikzpicture}

%% file: Figures/perf_rn50.tex
 \definecolor{bblue}{HTML}{4F81BD}
 \definecolor{rred}{HTML}{C0504D}
 \definecolor{ggreen}{HTML}{9BBB59}
 
\begin{tikzpicture}
\centering
\begin{axis}[
    title={ \textbf{ResNet-50}},
    width = 0.47\textwidth,
    height = 5.5 cm,
    legend pos=north west,
    xlabel=Accuracy (\%) - ImageNet,
    ylabel=Normalized energy cost ,
    xmin=74, xmax=76.4,
    ymin=0.38, ymax=1.22,
    xtick={73,73.5,74,74.5,75,75.5,76},
    enlargelimits=false,
    ymajorgrids=true,
    xmajorgrids=true,
     every tick label/.append style={font=\footnotesize},
     every axis label/.append style={font=\footnotesize},
   legend style={nodes={scale=0.6, transform shape}},
    grid style=dashed,
]

   \addplot[
    color=ggreen,
    mark=square*,
    mark size=1.5pt,
    line width=1.5
    ]
    coordinates {
    (76.1,1)(75.62,0.56)(75.50,0.53)(75.38,0.49)(75.06,0.44)(74.86,0.42)(74.68,0.4)(74.25,0.39)};
   \addlegendentry{RNS mixed}
    
    \addplot[
    color=bblue,
    mark=square*,
    mark size=1.5pt,
    line width=1.5
    ]
    coordinates {
    (76.1,1.17)(75.62,0.79)(75.51,0.61)(75.35,0.52)(75.16,0.45)(74.86,0.42)(74.63,0.40)(74.22,0.39)};
   \addlegendentry{FXP mixed}

    \addplot[
    only marks,
    color=rred,
    mark=diamond*,
    mark size=3pt
    ]
    coordinates {
    (76.1,1)(75.65,0.74)(72.71,0.46)}  node [left, pos=0.0] {W8A8}  node [left, pos=0.5] {W4A8}  node [right, pos=1.0] {W4A4};   
  \addlegendentry{RNS uniform }
  
  %
\addplot[thick, samples=50, smooth,domain=0:1,magenta, style=dashed] coordinates {(75.1,0)(75.1,1.3)} ;
\addplot[thick, samples=50, smooth,domain=0:1,black, style=dashed] coordinates {(76.13,0)(76.13,1.3)}  ;

\end{axis}
\end{tikzpicture}

%% file: Figures/perf_yolo5m.tex
 \definecolor{bblue}{HTML}{4F81BD}
 \definecolor{rred}{HTML}{C0504D}
 \definecolor{ggreen}{HTML}{9BBB59}
 
\begin{tikzpicture}
\centering
\begin{axis}[
    title={ \textbf{Yolov5-m}},
    width = 0.47\textwidth,
    height = 5.5 cm,
    legend pos=north west,
    xlabel=mAP (@0.5~IoU) - COCO,
    ylabel=Normalized energy cost ,
    xmin=60, xmax=64,
    ymin=0.5, ymax=1.22,
    enlargelimits=false,
    ymajorgrids=true,
    xmajorgrids=true,
     every tick label/.append style={font=\footnotesize},
     every axis label/.append style={font=\footnotesize},
   legend style={nodes={scale=0.6, transform shape}},
    grid style=dashed,
]

  \addplot[
    color=ggreen,
    mark=square*,
    mark size=1.5pt,
        line width=1.5
    ]
    coordinates {
    (63.7,1)(63.37,0.78)(63.19,0.70)(62.98,0.65)(62.54,0.61)(62.04,0.59)(61.68,0.57)(60.62,0.56)};
   \addlegendentry{RNS mixed}

    \addplot[
    color=bblue,
    mark=square*,
    mark size=1.5pt,
        line width=1.5
    ]
    coordinates {
    (63.7,1.2)(63.37,0.91)(63.25,0.82)(63.15,0.76)(62.77,0.69)(62.53,0.66)(62.13,0.63)(61.82,0.61)(61.08,0.59)(60.57,0.58)};
   \addlegendentry{FXP mixed}

   \addplot[
    only marks,
    color=rred,
    mark=diamond*,
    mark size=3pt
    ]
    coordinates {
    (63.7,1)(61.77,0.74)}  node [left, pos=0.0] {W8A8}  node [left, pos=0.5] {W4A8};
   
  \addlegendentry{RNS uniform }
  
   \addplot[thick, samples=50, smooth,domain=0:1,magenta, style=dashed] coordinates {(62.8,0)(62.8,1.27)}  ;
 \addplot[thick, samples=50, smooth,domain=0:1,black, style=dashed] coordinates {(63.8,0)(63.8,1.27)};

\end{axis}
\end{tikzpicture}

%% file: Figures/perf_bert.tex
 \definecolor{bblue}{HTML}{4F81BD}
 \definecolor{rred}{HTML}{C0504D}
 \definecolor{ggreen}{HTML}{9BBB59}
 
\begin{tikzpicture}
\centering
\begin{axis}[
    title={ \textbf{Bert-base}},
    width = 0.47\textwidth,
    height = 5.5 cm,
    legend pos=north west,
    xlabel=F1 Score - SQuAD,
    ylabel=Normalized energy cost ,
    xmin=80.2, xmax=89,
    ymin=0.45, ymax=1.22,
    enlargelimits=false,
    ymajorgrids=true,
    xmajorgrids=true,
     every tick label/.append style={font=\small},
     every axis label/.append style={font=\small},
   legend style={nodes={scale=0.6, transform shape}},
    grid style=dashed,
]

 \addplot[
    color=ggreen,
    mark=square*,
    mark size=1.5pt,
    line width=1.5
    ]
    coordinates {
    (87.86,1)
    (87.34,0.73)
    (86.89,0.65)
    (86.64,0.61)
    (86.27,0.59)
    (85.72,0.56)
    (85.21,0.54)
    (84.73,0.52)
    (82.16,0.51)
    (80.26,0.50)
    };
   \addlegendentry{RNS mixed}

    \addplot[
    color=bblue,
    mark=square*,
    mark size=1.5pt,
        line width=1.5
    ]
    coordinates {
    (87.86,1.2)  
    (87.43,0.84)
    (86.85,0.72)
    (86.63,0.66)
    (86.37,0.63)
    (85.66,0.60)
    (85.42,0.58)
    (84.92,0.56)
    (82.47,0.55)
    (80.64,0.54)
    };
   \addlegendentry{BNS mixed}

    \addplot[
    only marks,
    color=rred,
    mark=diamond*,
    mark size=3pt
    ]
    coordinates {
    (87.87,1)(87.26,0.74) }  node [above, pos=0.0,xshift=-4pt] {W8A8} node [right, pos=1.0] {W4A8};
\addlegendentry{RNS uniform}

   \addplot[thick, samples=50, smooth,domain=0:1,magenta, style=dashed] coordinates {(87.50,0)(87.50,1.27)}  ;
 \addplot[thick, samples=50, smooth,domain=0:1,black, style=dashed] coordinates {(88.54,0)(88.54,1.27)};

\end{axis}
\end{tikzpicture}

%% file: Figures/perf_vit16.tex
 \definecolor{bblue}{HTML}{4F81BD}
 \definecolor{rred}{HTML}{C0504D}
 \definecolor{ggreen}{HTML}{9BBB59}

\begin{tikzpicture}
\centering
\begin{axis}[
    title={ \textbf{ViT-b16}},
    width = 0.47\textwidth,
    height = 5.5 cm,
    legend pos=north west,
    xlabel=Accuracy (\%) - ImageNet,
    ylabel=Normalized energy cost ,
    xmin=79, xmax=82,
    ymin=0.40, ymax=1.22,
    enlargelimits=false,
    ymajorgrids=true,
    xmajorgrids=true,
     every tick label/.append style={font=\footnotesize},
     every axis label/.append style={font=\footnotesize},
   legend style={nodes={scale=0.6, transform shape}},
    grid style=dashed,
]

    \addplot[
    color=ggreen,
    mark=square*,
    mark size=1.5,
        line width=1.5
    ]
    coordinates {
    (81.44,1)(80.75,0.67)(80.44,0.59)(80.37,0.57)(80.30,0.55)(80.20,0.51)(79.90,0.48)(79.07,0.43)} ;
   \addlegendentry{RNS mixed}

   \addplot[
    color=bblue,
    mark=square*,
    mark size=1.5,
    line width=1.5
    ]
    coordinates {
    (81.44,1.20)(80.75,0.80)(80.44,0.68)(80.36,0.64)(80.30,0.61)(80.18,0.55)(79.26,0.46)} ;
   \addlegendentry{FXP mixed}

\addplot[
    only marks,
    color=rred,
    mark=diamond*,
    mark size=3pt
    ]
    coordinates {
    (81.44,1) (80.92,0.75)}   node [above, pos=0.0] {W8A8}  node [left, pos=0.5] {W4A8}  ;  
  \addlegendentry{RNS-uniform}

   \addplot[thick, samples=50, smooth,domain=0:1,magenta, style=dashed] coordinates {(80.44,0)(80.44,1.3)} ;
      \addplot[thick, samples=50, smooth,domain=0:1,black, style=dashed] coordinates {(81.44,0)(81.44,1.3)}  ;

\end{axis}
\end{tikzpicture}

%% file: references.bib
@article{obc,
  title={{Optimal brain compression: A framework for accurate post-training quantization and pruning}},
  author={Frantar, Elias and Alistarh, Dan},
  journal={Advances in Neural Information Processing Systems},
  volume={35},
  pages={4475--4488},
  year={2022}
}

@article{mxp1,
  title={{Improving post training neural quantization: Layer-wise calibration and integer programming}},
  author={Hubara, Itay and Nahshan, Yury and Hanani, Yair and Banner, Ron and Soudry, Daniel},
  journal={arXiv preprint arXiv:2006.10518},
  year={2020}
}

@article{sakellariou2024,
   author={Sakellariou, Vasilis and Paliouras, Vassilis and Kouretas, Ioannis and Saleh, Hani and Stouraitis, Thanos},
  journal={IEEE Transactions on Emerging Topics in Computing}, 
  title={A 22-nm 4.92 TOPS/W end-to-end RNS DNN Accelerator for Edge-AI Devices}, 
  year={2026},
  volume={},
  number={},
  pages={1-16},
  doi={10.1109/TETC.2026.3689523}}

@INPROCEEDINGS{diana,
  author={Ueyoshi, Kodai and Papistas, Ioannis A. and Houshmand, Pouya and Sarda, Giuseppe M. and Jain, Vikram and Shi, Man and Zheng, Qilin and Giraldo, Sebastian and Vrancx, Peter and Doevenspeck, Jonas and Bhattacharjee, Debjyoti and Cosemans, Stefan and Mallik, Arindam and Debacker, Peter and Verkest, Diederik and Verhelst, Marian},
  booktitle={2022 IEEE International Solid-State Circuits Conference (ISSCC)}, 
  title={{DIANA: An End-to-End Energy-Efficient Digital and ANAlog Hybrid Neural Network SoC}}, 
  year={2022},
  volume={65},
  number={},
  pages={1-3},
  doi={10.1109/ISSCC42614.2022.9731716}}

@ARTICLE{bitblade,
  author={Ryu, Sungju and Kim, Hyungjun and Yi, Wooseok and Kim, Eunhwan and Kim, Yulhwa and Kim, Taesu and Kim, Jae-Joon},
  journal={IEEE Journal of Solid-State Circuits}, 
  title={{BitBlade: Energy-Efficient Variable Bit-Precision Hardware Accelerator for Quantized Neural Networks}}, 
  year={2022},
  volume={57},
  number={6},
  pages={1924-1935},
  doi={10.1109/JSSC.2022.3141050}}

@INPROCEEDINGS{marsellus,
  author={Conti, Francesco and Rossi, Davide and Paulin, Gianna and Garofalo, Angelo and Di Mauro, Alfio and Rutishauer, Georg and Ottavi, Gian marco and Eggimann, Manuel and Okuhara, Hayate and Huard, Vincent and Montfort, Olivier and Jure, Lionel and Exibard, Nils and Gouedo, Pascal and Louvat, Mathieu and Botte, Emmanuel and Benini, Luca},
  booktitle={2023 IEEE International Solid-State Circuits Conference (ISSCC)}, 
  title={{22.1 A 12.4TOPS/W @ 136GOPS AI-IoT System-on-Chip with 16 RISC-V, 2-to-8b Precision-Scalable DNN Acceleration and 30\%-Boost Adaptive Body Biasing}}, 
  year={2023},
  volume={},
  number={},
  pages={21-23},
  doi={10.1109/ISSCC42615.2023.10067643}}

@inproceedings{bert,
  title={{BERT: Pre-training of Deep Bidirectional Transformers for Language Understanding}},
  author={Jacob Devlin and Ming-Wei Chang and Kenton Lee and Kristina Toutanova},
  booktitle={North American Chapter of the Association for Computational Linguistics},
  year={2019},
  url={https://api.semanticscholar.org/CorpusID:52967399}
}

@INPROCEEDINGS{bitfusion,

  author={Sharma, Hardik and Park, Jongse and Suda, Naveen and Lai, Liangzhen and Chau, Benson and Kim, Joon Kyung and Chandra, Vikas and Esmaeilzadeh, Hadi},

  booktitle={2018 ACM/IEEE 45th Annual International Symposium on Computer Architecture (ISCA)}, 

  title={{Bit Fusion: Bit-Level Dynamically Composable Architecture for Accelerating Deep Neural Network}}, 

  year={2018},

  volume={},

  number={},

  pages={764-775},

  doi={10.1109/ISCA.2018.00069}}

@article{yolo,
  title={{Yolov3: An incremental improvement}},
  author={Redmon, Joseph and Farhadi, Ali},
  journal={arXiv preprint arXiv:1804.02767},
  year={2018}
}

@INPROCEEDINGS{resnet,
  author={He, Kaiming and Zhang, Xiangyu and Ren, Shaoqing and Sun, Jian},
  booktitle={2016 IEEE Conference on Computer Vision and Pattern Recognition (CVPR)}, 
  title={{Deep Residual Learning for Image Recognition}}, 
  year={2016},
  volume={},
  number={},
  pages={770-778},
  doi={10.1109/CVPR.2016.90}}

@INPROCEEDINGS{rns5,
  author={Roohi, Arman and Taheri, MohammadReza and Angizi, Shaahin and Fan, Deliang},
  booktitle={2021 IEEE/ACM International Conference On Computer Aided Design (ICCAD)}, 
  title={{RNSiM: Efficient Deep Neural Network Accelerator Using Residue Number Systems}}, 
  year={2021},
  volume={},
  number={},
  pages={1-9},
  doi={10.1109/ICCAD51958.2021.9643531}}

@article{rns4,
title = {{Application of the residue number system to reduce hardware costs of the convolutional neural network implementation}},
journal = {{Mathematics and Computers in Simulation}},
volume = {177},
pages = {232-243},
year = {2020},
issn = {0378-4754},
doi = {https://doi.org/10.1016/j.matcom.2020.04.031},
author = {M.V. Valueva and N.N. Nagornov and P.A. Lyakhov and G.V. Valuev and N.I. Chervyakov}}

@ARTICLE{rns2,  
  author={N. {Samimi} and M. {Kamal} and A. {Afzali-Kusha} and M. {Pedram}}, 
  journal={{IEEE Transactions on Circuits and Systems I: Regular Papers}},   
  title={{Res-DNN: A Residue Number System-Based DNN Accelerator Unit}},   
  year={2020},  
  volume={67},  
  number={2},  
  pages={658-671},  
  doi={10.1109/TCSI.2019.2951083}}

@article{dim1mult,
author = {Efstathiou, C. and Vergos, Haridimos and Dimitrakopoulos, Giorgos and Nikolos, Dimitris},
year = {2005},
month = {04},
pages = {491-496},
title = {{Efficient Diminished1 Modulo $2^n+1$ Multipliers}},
volume = {54},
journal = {IEEE Transactions on Computers - TC}
}

@INPROCEEDINGS{rns3,  author={H. {Nakahara} and T. {Sasao}},  booktitle={{2015 25th International Conference on Field Programmable Logic and Applications (FPL)}},   title={A deep convolutional neural network based on nested residue number system},   year={2015},  volume={},  number={},  pages={1-6},  doi={10.1109/FPL.2015.7293933}}

@inproceedings{rns6,
author = {Salamat, Sahand and Imani, Mohsen and Gupta, Saransh and Rosing, Tajana},
year = {2018},
month = {11},
pages = {1-12},
title = {{RNSnet: In-Memory Neural Network Acceleration Using Residue Number System}},
doi = {10.1109/ICRC.2018.8638592}}

@article{rns1,
  title={{A 22-nm 4.92 TOPS/W end-to-end RNS DNN Accelerator for Edge-AI Devices}},
  author={Sakellariou, Vasilis and Paliouras, Vassilis and Kouretas, Ioannis and Saleh, Hani and Stouraitis, Thanos},
journal={Authorea Preprints},
year = {2024},
}

@article{rns8,
  title={{A multiplier-Free RNS-Based CNN accelerator exploiting bit-Level sparsity}},
  author={Sakellariou, Vasilis and Paliouras, Vassilis and Kouretas, Ioannis and Saleh, Hani and Stouraitis, Thanos},
  journal={IEEE Transactions on Emerging Topics in Computing},
  pages={1--16},
  year={2023},
  publisher={IEEE Computer Society}
}

@INPROCEEDINGS{dnnaccel1,
  author={Park, Jun-Seok and Park, Changsoo and Kwon, Suknam and Kim, Hyeong-Seok and Jeon, Taeho and Kang, Yesung and Lee, Heonsoo and Lee, Dongwoo and Kim, James and Lee, YoungJong and Park, Sangkyu and Jang, Jun-Woo and Ha, SangHyuck and Kim, MinSeong and Bang, Jihoon and Lim, Suk Hwan and Kang, Inyup},
  booktitle={2022 IEEE International Solid-State Circuits Conference (ISSCC)}, 
  title={A Multi-Mode 8K-MAC HW-Utilization-Aware Neural Processing Unit with a Unified Multi-Precision Datapath in 4nm Flagship Mobile SoC}, 
  year={2022},
  volume={65},
  number={},
  pages={246-248},
  doi={10.1109/ISSCC42614.2022.9731639}}

@ARTICLE{dnnaccel2,
  author={Mo, Huiyu and Zhu, Wenping and Hu, Wenjing and Li, Qiang and Li, Ang and Yin, Shouyi and Wei, Shaojun and Liu, Leibo},
  journal={IEEE Journal of Solid-State Circuits}, 
  title={A 12.1 TOPS/W Quantized Network Acceleration Processor With Effective-Weight-Based Convolution and Error-Compensation-Based Prediction}, 
  year={2022},
  volume={57},
  number={5},
  pages={1542-1557},
  doi={10.1109/JSSC.2021.3113569}}

@INPROCEEDINGS{dnnaccel3,
  author={Dong, Pingcheng and Tan, Yonghao and Liu, Xuejiao and Luo, Peng and Liu, Yu and Liang, Luhong and Zhou, Yitong and Pang, Di and Yung, Man-To and Zhang, Dong and Huang, Xijie and Liu, Shih-Yang and Wu, Yongkun and Tian, Fengshi and Tsui, Chi-Ying and Tu, Fengbin and Cheng, Kwang-Ting},
  booktitle={2025 IEEE International Solid-State Circuits Conference (ISSCC)}, 
  title={A 28nm 0.22uJ/Token Memory-Compute-Intensity-Aware CNN-Transformer Accelerator with Hybrid-Attention-Based Layer-Fusion and Cascaded Pruning for Semantic-Segmentation}, 
  year={2025},
  volume={68},
  number={},
  pages={01-03},
  doi={10.1109/ISSCC49661.2025.10904499}}

@ARTICLE{dnnaccel4,
  author={Han, Donghyeon and Chandrakasan, Anantha P.},
  journal={IEEE Journal of Solid-State Circuits}, 
  title={MEGA.mini: An Energy-Efficient NPU Leveraging a Novel Big/Little Core With Hybrid Input Activation for Generative AI Acceleration}, 
  year={2025},
  volume={},
  number={},
  pages={1-14},
  doi={10.1109/JSSC.2025.3626894}}

@INPROCEEDINGS{dnnaccel5,
  author={Moon, Seunghyun and Li, Mao and Chen, Gregory K. and Knag, Phil C. and Krishnamurthy, Ram Kumar and Seok, Mingoo},
  booktitle={2025 IEEE International Solid-State Circuits Conference (ISSCC)}, 
  title={T-REX: A 68-to-567us/Token 0.41-to-3.95uJ/Token Transformer Accelerator with Reduced External Memory Access and Enhanced Hardware Utilization in 16nm FinFET}, 
  year={2025},
  volume={68},
  number={},
  pages={406-408},
  doi={10.1109/ISSCC49661.2025.10904793}}

@INPROCEEDINGS{dnnaccel6,
  author={Du, Cheng-Yan and Tsai, Chieh-Fu and Chen, Wen-Ching and Lin, Liang-Yi and Chang, Nian-Shyang and Lin, Chun-Pin and Chen, Chi-Shi and Yang, Chia-Hsiang},
  booktitle={2023 IEEE International Solid-State Circuits Conference (ISSCC)}, 
  title={A 28nm 11.2TOPS/W Hardware-Utilization-Aware Neural-Network Accelerator with Dynamic Dataflow}, 
  year={2023},
  volume={},
  number={},
  pages={1-3},
  doi={10.1109/ISSCC42615.2023.10067774}}

@ARTICLE{jssc25,
  author={Verhelst, Marian and Benini, Luca and Verma, Naveen},
  journal={IEEE Journal of Solid-State Circuits}, 
  title={How to Keep Pushing ML Accelerator Performance? Know Your Rooflines!}, 
  year={2025},
  volume={60},
  number={6},
  pages={1888-1905},
  doi={10.1109/JSSC.2025.3553765}}

@ARTICLE{xpulpnn,
  author={Garofalo, Angelo and Tagliavini, Giuseppe and Conti, Francesco and Benini, Luca and Rossi, Davide},
  journal={IEEE Transactions on Emerging Topics in Computing}, 
  title={XpulpNN: Enabling Energy Efficient and Flexible Inference of Quantized Neural Networks on RISC-V Based IoT End Nodes}, 
  year={2021},
  volume={9},
  number={3},
  pages={1489-1505},
  doi={10.1109/TETC.2021.3072337}}

@ARTICLE{vega,
  author={Rossi, Davide and Conti, Francesco and Eggiman, Manuel and Mauro, Alfio Di and Tagliavini, Giuseppe and Mach, Stefan and Guermandi, Marco and Pullini, Antonio and Loi, Igor and Chen, Jie and Flamand, Eric and Benini, Luca},
  journal={IEEE Journal of Solid-State Circuits}, 
  title={Vega: A Ten-Core SoC for IoT Endnodes With DNN Acceleration and Cognitive Wake-Up From MRAM-Based State-Retentive Sleep Mode}, 
  year={2022},
  volume={57},
  number={1},
  pages={127-139},
  doi={10.1109/JSSC.2021.3114881}}

@INPROCEEDINGS{mx,
  author={Perotti, Matteo and Zhang, Yichao and Cavalcante, Matheus and Mustafa, Enis and Benini, Luca},
  booktitle={2024 Design, Automation and Test in Europe Conference and Exhibition (DATE)}, 
  title={MX: Enhancing RISC-V's Vector ISA for Ultra-Low Overhead, Energy-Efficient Matrix Multiplication}, 
  year={2024},
  volume={},
  number={},
  pages={1-6},
  doi={10.23919/DATE58400.2024.10546720}}

@INPROCEEDINGS{flexv,
  author={Nadalini, Alessandro and Rutishauser, Georg and Burrello, Alessio and Bruschi, Nazareno and Garofalo, Angelo and Benini, Luca and Conti, Francesco and Rossi, Davide},
  booktitle={2023 IEEE Computer Society Annual Symposium on VLSI (ISVLSI)}, 
  title={A 3 TOPS/W RISC-V Parallel Cluster for Inference of Fine-Grain Mixed-Precision Quantized Neural Networks}, 
  year={2023},
  volume={},
  number={},
  pages={1-6},
  doi={10.1109/ISVLSI59464.2023.10238679}}

@INPROCEEDINGS{datamaestro,
  author={Yi, Xiaoling and Deng, Yunhao and Antonio, Ryan and Kong, Fanchen and Paim, Guilherme and Verhelst, Marian},
  booktitle={2025 62nd ACM/IEEE Design Automation Conference (DAC)}, 
  title={DataMaestro: A Versatile and Efficient Data Streaming Engine Bringing Decoupled Memory Access To Dataflow Accelerators}, 
  year={2025},
  volume={},
  number={},
  pages={1-7},
  doi={10.1109/DAC63849.2025.11133141}}

@inproceedings{quadrilatero,
  title={Quadrilatero: A RISC-V programmable matrix coprocessor for low-power edge applications},
  author={Cammarata, Danilo and Perotti, Matteo and Bertuletti, Marco and Garofalo, Angelo and Schiavone, Pasquale Davide and Atienza, David and Benini, Luca},
  booktitle={Proceedings of the 22nd ACM International Conference on Computing Frontiers: Workshops and Special Sessions},
  pages={66--69},
  year={2025}
}

@article{spatz,
  title={Spatz: Clustering compact risc-v-based vector units to maximize computing efficiency},
  author={Cavalcante, Matheus and Perotti, Matteo and Riedel, Samuel and Benini, Luca},
  journal={arXiv preprint arXiv:2309.10137},
  year={2023}
}

@INPROCEEDINGS{iscas2025sakel,
  author={Sakellariou, Vasilis and Paliouras, Vassilis and Kouretas, Ioannis and Saleh, Hani and Stouraitis, Thanos},
  booktitle={2025 IEEE International Symposium on Circuits and Systems (ISCAS)}, 
  title={A Mixed-Precision RNS DNN Accelerator}, 
  year={2025},
  volume={},
  number={},
  pages={1-5},
  doi={10.1109/ISCAS56072.2025.11043322}}

@ARTICLE{nnsurvey,
  author={Alsuhli, Ghada and Sakellariou, Vasilis and Saleh, Hani and Al-Qutayri, Mahmoud and Mohammad, Baker and Stouraitis, Thanos},
  journal={Proceedings of the IEEE}, 
  title={A Survey and Comparative Analysis of Number Systems for Deep Neural Networks}, 
  year={2025},
  volume={113},
  number={2},
  pages={172-207},
  doi={10.1109/JPROC.2025.3578756}}

@ARTICLE{riscvattent,
  author={Wiese, Philip and İslamoğlu, Gamze and Scherer, Moritz and Macan, Luka and Jung, Victor Jean-Baptiste and Burrello, Alessio and Conti, Francesco and Benini, Luca},
  journal={IEEE Design \& Test}, 
  title={Toward Attention-Based TinyML: A Heterogeneous Accelerated Architecture and Automated Deployment Flow}, 
  year={2025},
  volume={42},
  number={5},
  pages={63-72},
  doi={10.1109/MDAT.2025.3527371}}

@article{onlinenorm,
  title={Online normalizer calculation for softmax},
  author={Milakov, Maxim and Gimelshein, Natalia},
  journal={arXiv preprint arXiv:1805.02867},
  year={2018}
}

@article{vit,
  title={An image is worth 16x16 words: Transformers for image recognition at scale},
  author={Dosovitskiy, Alexey and Beyer, Lucas and Kolesnikov, Alexander and Weissenborn, Dirk and Zhai, Xiaohua and Unterthiner, Thomas and Dehghani, Mostafa and Minderer, Matthias and Heigold, Georg and Gelly, Sylvain and others},
  journal={arXiv preprint arXiv:2010.11929},
  year={2020}
}

@article{mobilenet,
  title={Mobilenets: Efficient convolutional neural networks for mobile vision applications},
  author={Howard, Andrew G and Zhu, Menglong and Chen, Bo and Kalenichenko, Dmitry and Wang, Weijun and Weyand, Tobias and Andreetto, Marco and Adam, Hartwig},
  journal={arXiv preprint arXiv:1704.04861},
  year={2017}
}
